\documentclass[twocolumn,showpacs,preprintnumbers,amsmath,amssymb,nofootinbib]{revtex4-2}
\usepackage{epsfig}

\usepackage{comment}
\usepackage[table]{xcolor}
\usepackage{graphicx}
\usepackage{dcolumn}
\usepackage{bm}

\usepackage{hyperref}
\hypersetup{colorlinks=true,linkcolor=magenta,anchorcolor=green,citecolor=cyan,filecolor=black,menucolor=black,urlcolor=brown}
\usepackage{slashed}
\newcommand{\ba}{\begin{eqnarray}}
\newcommand{\ea}{\end{eqnarray}}
\newcommand{\be}{\begin{equation}}
\newcommand{\ee}{\end{equation}}

\begin{document}

\title{Repulsive ultralight dark matter from Hosotani mechanism}

\author{Philippe Brax}
\affiliation{Universit\'{e} Paris-Saclay, CNRS, CEA, Institut de physique th\'{e}orique, 91191, Gif-sur-Yvette, France}
\author{Patrick Valageas}
\affiliation{Universit\'{e} Paris-Saclay, CNRS, CEA, Institut de physique th\'{e}orique, 91191, Gif-sur-Yvette, France}

\begin{abstract}

We build an ultralight dark matter model with repulsive self-interactions, starting from a 5D action
that only includes a $U(1)$ gauge field and several massive and charged free fermions. 
The dark matter scalar field corresponds to the fifth component of the gauge field,
after compactification to 4D.
Although the single-fermion case only leads to attractive self-interactions, two fermions can already
give rise to repulsive self-interactions, thanks to a Scherk-Schwarz twist around the fifth dimension.
We check the observational and theoretical self-consistency of this scenario, from the inflation 
stage to the current time. We find that large ranges of model parameters are allowed.
For scalar masses below $10^{-12}$ eV the self-interactions are negligible and the model behaves
as fuzzy dark matter. For higher masses the repulsive self-interactions govern the formation
of solitons of astrophysical size. Larger sizes require a large initial misalignment
or small masses in the fuzzy dark matter regime.

\end{abstract}
\maketitle

\section{Introduction}

Ultralight dark matter (ULDM), with a particle mass $m \lesssim 1$ eV, is a viable alternative to the 
WIMP scenario \cite{Hui:2016ltb}. In these models, the core of dark matter halos can be
altered compared to the Cold Dark Matter (CDM) behavior, because of wavelike effects
that come into play on the large de Broglie scale $\lambda_{\rm dB} = 2\pi/(mv)$ or because
of self-interactions \cite{Goodman:2000tg,Hu2000,Chavanis:2011zi} . 
This can lead to the formation of solitons (i.e. hydrostatic equilibria) with flat cores at the center
of dark matter halos or galaxies \cite{Kawai:2023okm}. This may alleviate
some of the short distance issues of the standard model of cosmology without the need for baryonic 
back-reaction effects \cite{DelPopolo:2016emo}. 
In the absence of self-interactions, i.e. the fuzzy dark matter (FDM) case, solitons can reach
$kpc$ size for $m \sim 10^{-22}$ eV. However, such low masses are ruled out by observations
of the Lyman-$\alpha$ forest \cite{Armengaud:2017nkf}.
If the self-interactions are repulsive and strong enough, one could obtain large soliton sizes
for masses in the full ULDM range $10^{-21} \, {\rm eV} < m < 1 \, {\rm eV}$.
Another difference between the FDM and self-interacting cases are the different
power-law scalings between the mass and radius of the solitons \cite{Brax:2019fzb}.
If the solitons are of much smaller astrophysical size, they are also called boson stars.
This also provides a satisfactory dark matter model, which behaves like CDM on galactic and
 cosmological scales and can only be distinguished through its effects on astrophysical
 scales, for instance on the gravitational waves emitted by binary systems embedded in 
 such dark matter clouds \cite{Brax:2024yqh}. 
 
Attractive scalar models are natural and appear from pseudo-Goldstone modes with 
a trigonometric potential like in Quantum Chromo Dynamics (QCD) 
\cite{Wilczek:1977pj,Weinberg:1977ma,Preskill:1982cy}. They have a negative quartic 
self-interaction. In this paper, we investigate the opposite situation with positive self-couplings 
and repulsive self-interactions. 
This is a less common configuration and typically requires a tuning between the couplings
of several fermion and boson fields \cite{Fan2016}.
We present in this paper a somewhat more economical model, with only one $U(1)$ gauge field 
and two fermions.
This is achieved by extending the construction of axion models from 5D by considering the dark 
matter field as the fifth component of a 5D gauge field \cite{Choi:2003wr}. The coupling of fermions 
to the gauge field leads to a one-loop potential for the dark matter field that can be
repulsive as soon as the fermions are massive and charged. They also must have a non-trivial 
boundary condition around the fifth dimension with a Scherk-Schwarz twist \cite{Scherk:1979zr}. 
Within this framework, we find that repulsive interactions are not exceptional but form 
a large part of the parameter space.
This approach provides repulsive quartic self-interactions for ultralight dark matter
particles contrary to the non-perturbative axion potential or supersymmetric models
\cite{DIVECCHIA1980253,di_Cortona_2016,COLEMAN1985263,Kusenko:1997si,Kasuya:2025ubo}, which are generically attractive.
 
We focus on pre-inflationary scenarios and one advantage of this explicit model is that
it allows us to check in detail its consistency with observational and theoretical constraints,
from the inflation era to current times.
We will find that a large range of model parameters is allowed, with masses covering
the full range $10^{-21} \, {\rm eV} < m < 1 \, {\rm eV}$.
However, the solitons are of astrophysical rather than galactic size, unless the initial 
misalignment is very large.
The scale of the fifth dimension is chosen so as to recover the measured dark matter 
cosmological mean density today. For generic misalignment angles $\theta_i \sim 1$ and
fermion masses $m_\psi R \lesssim 1$, we find in Eq.(\ref{eq:R-m}) that the mass range 
$10^{-21} \, {\rm eV} < m < 1 \, {\rm eV}$ gives
$2 \, {\rm keV} \lesssim 1/R \lesssim 200 \, {\rm GeV}$, i.e. a typical radius 
$10^{-18} \, {\rm m} \lesssim R \lesssim 10^{-10} \, {\rm m}$,
far below the current sensitivity of tests of gravity ($R \lesssim 10^{-5}$ m) and well above 
the inflationary scale $H_I$ as required by self-consistency and observational constraints.

The paper is arranged as follows. 
In Section~\ref{sec:spacetime}, we describe the 5D action of our model, which includes one
$U(1)$ gauge field and $N_f$ free fermions. We compactify to 4D and integrate over the
Kaluza-Klein towers of dark photons and fermions to derive the potential of the ultralight
dark matter scalar field.
In Section~\ref{sec:building-repulsive}, we show that it is possible to obtain a repulsive 
self-interaction by including two fermions.
In Section~\ref{sec:late-time-constraints} we check the consistency of the model with 
late-time observations, such as the abundance of dark matter and small-scale tests of gravity.
In Section~\ref{sec:pre-inflationary} we verify the conditions associated with the validity
of the model during the inflation era, such as the size of the fifth dimension and the
gravitational production of dark fermions and photons.
In Section~\ref{sec:theoretical-consistency}, we check the theoretical self-consistency
of our derivation of the dark matter potential: the validity of our derivative expansion
and the allowed range for the UV cutoff of the 5D theory.
In Section~\ref{sec:parameter-space} we obtain the parameter space of the model that is consistent
with all these conditions, in terms of the scalar mass $m$, initial misalignment angle or soliton
size.
In Section~\ref{sec:low-mass} we consider in more details the case of low-mass fermions,
where the minimum of the potential with repulsive self-interactions is only a local minimum.
We check that its lifetime is so large that for practical purposes it can be considered as a legitimate
minimum, as the true minimum obtained for large-mass fermions.
We conclude in Section~\ref{sec:Conclusion}.

Several Appendices detail our computations, for the integration over the fermion Kaluza-Klein
towers in App.~\ref{app:KK-modes}, the validity of the derivative expansion in
App.~\ref{app:derivative-expansion}, the gravitational productions of dark fermions in
App.~\ref{app:production-fermions} and of dark photons in App.~\ref{app:production-photons}.

\section{Spacetime and Field Content}
\label{sec:spacetime}

\subsection{Compactification from 5D to 4D}
\label{sec:compactification}

We consider a 5D U(1) gauge theory with the spacetime 
$\mathcal{M}_5 = \mathbb{R}^{1,3} \times S^1$, where the fifth dimension $y$ is compactified 
on the circle $S^1$ of radius $R$. 
The field content includes a 5D U(1) gauge field $A_M = (A_\mu, A_5)$ and $N_f$ 5D Dirac fermions 
$\psi_j$, charged under U(1).
The 5D Lagrangian is given by:
\ba
\mathcal{L}_{5D}[ \psi_j,A_M] & \! = \! & \sum_{j=1}^{N_f} \left[ i \bar{\psi}_j \Gamma^M 
(\partial_M \! + \! i g_5 q_j A_M) \psi_j - m_j \bar{\psi}_j \psi_j \right] \nonumber \\
&& -\frac{1}{4} F_{MN} F^{MN} ,
\label{eq:L-5D}
\ea
where $F_{MN} = \partial_M A_N - \partial_N A_M$ is the field strength tensor,
$q_j$ and $m_j$ are the fermion charges and masses.
We work in mostly-minus signature $\eta_{MN} = \text{diag}(+1,-1,-1,-1,-1)$, for
$M,N=0,1,2,3,5$. The 5D matrices $\Gamma^M$ are constructed from the usual 4D Dirac matrices
\be
\Gamma^\mu = \gamma^\mu, \qquad \Gamma^5 = i\gamma^5, \qquad \mu=0,1,2,3,
\ee
where $\gamma^5 = i\gamma^0\gamma^1\gamma^2\gamma^3$, and we have
$\{\Gamma^M,\Gamma^N\} = 2\eta^{MN}$.

The Lagrangian (\ref{eq:L-5D}) obeys the 5D gauge invariance
$A_M \to A_M - \frac{1}{g_5} \partial_M \alpha$, $\psi_j \to e^{i q_j \alpha} \psi_j$.
Thanks to this 5D gauge invariance, we can choose a gauge where $A_5(x,y)=\phi(x)/\sqrt{2\pi R}$
is independent of $y$ \cite{sundrum2005} and $\phi(x)$ will be the dark matter ``axion'' of the 4D model\footnote{5D gauge invariance requires that $A_5 \to A_5 - \frac{1}{g_5} \partial_5 \alpha$. Fixing $\phi(x)$ implies that $\alpha(x, y)\equiv \alpha (x)$, i.e. this gauge choice restricts gauge invariance to a residual symmetry corresponding to 4D gauge invariance.} .
The boundary conditions for the fields are:
\ba
&& A_\mu(x, y + 2\pi R) = A_\mu(x, y) , \nonumber \\
&& \psi_j(x, y+ 2\pi R) = e^{i 2\pi  \alpha_j} \psi_j(x, y) , \;\;\; 0 \leq \alpha_j < 1 ,
\ea
where for the fermions $\psi_j$ we introduced a Scherk-Schwarz twist $\alpha_j$
\cite{SCHERK197960,SCHERK197961,quiros2003,csaki2004}.
Due to these periodicity conditions, the fields are expanded as
\ba
A_\mu(x,y) & = & \frac{A_\mu^{(0)}(x)}{\sqrt{2\pi R}} + \sum_{n=1}^{\infty} \frac{1}{\sqrt{\pi R}} 
\big[A_\mu^{(c,n)}(x)\cos(ny/R) \nonumber \\
&& + A_\mu^{(s,n)}(x)\sin(ny/R)\big] , \nonumber \\
\psi_j(x, y) & = & \sum_{n=-\infty}^{\infty} \frac{\psi^{(n)}_j(x)}{\sqrt{2\pi R}} e^{i(n+\alpha_j)y/R}. 
\ea
After $y$-integration over $[0,2\pi R[$, we obtain
\ba
&& \mathcal{L}_{4D} \left[ \psi_j^{(n)},\phi,A_\mu^{(0)},A_\mu^{(c,n)},A_\mu^{(s,n)} \right] = 
\frac{1}{2} \partial_\mu \phi \partial^\mu \phi \nonumber \\
&& - \frac{1}{4} F_{\mu\nu}^{(0)} F^{(0)\mu\nu} + \sum_{n=1}^{\infty} \mathcal{L}_F^{(n)} \left[ A_\mu^{(c,n)} 
\right] +  \sum_{n=1}^{\infty} \mathcal{L}_F^{(n)} \left[ A_\mu^{(s,n)} \right] \nonumber \\
&& + \sum_{j=1}^{N_f} \sum_{n=-\infty}^{\infty} \mathcal{L}_{\psi_j,A}^{(n)} \left[ {\psi_j^{(n)}} \right] ,
\label{eq:L-4D-tot}
\ea
with
\be
n \geq 1 : \;\; \mathcal{L}_F^{(n)} \left[ A_\mu \right] = - \frac{1}{4} F_{\mu\nu} F^{\mu\nu}
+ \frac{n^2}{2 R^2} A_{\mu} A^{\mu} ,
\label{eq:L-Amu}
\ee
and
\be
\mathcal{L}_{\psi_j,A}^{(n)}[\psi] =  \bar{\psi} \left[ i \gamma^\mu \partial_\mu - m^A_j 
- i \zeta_{j,n} \gamma^5 \right] \psi ,
\label{eq:L-psi-j-n-A}
\ee
where we introduced the effective fermion masses
\ba
&& m_j^A = m_j + \frac{q_j}{2\pi R f} \gamma^\mu A_\mu^{(0)} , \nonumber \\
&& \zeta_{j,n} =  \frac{n + \alpha_j + q_j \phi/(2\pi f)}{R} ,
\label{eq:zeta-def}
\ea
and the ``decay constant''
\be
f = \frac{1}{g_5 \sqrt{2\pi R}} .
\label{eq:f-def}
\ee

\subsection{Integration over dark photons and fermions}
\label{sec:integration-photons-fermions}

The fields $A_\mu^{(n,c)}$ and $A_\mu^{(n,s)}$ have acquired a mass
$m_n=n/R$ and are governed by the Proca Lagrangian of massive spin-1 fields \cite{Schwartz_2013}. 
They are decoupled from the other fields. Therefore, they can be integrated over
and have no impact on the dynamics, except through gravitational effects.
We shall check in Section~\ref{sec:Photon-prod} below that they are not gravitationally produced
during the inflation era, because of their very large mass. This is guaranteed as soon as
$1/R \gg H_I$, where $H_I$ is the inflation Hubble rate.
This condition is verified in our model and it is also required to recover a standard 4D inflationary
stage in Eq.(\ref{eq:R-HI}) below.
After integration of the massive spin-1 fields $A_\mu^{(n,c)}$ and $A_\mu^{(n,s)}$, we are left
with the Lagrangian
\ba
\mathcal{L}_{4D} \! \left[ \psi_j^{(n)},\phi,A_\mu^{(0)} \right] \! & = & \!
\frac{1}{2} \partial_\mu \phi \partial^\mu \phi  - \frac{1}{4} F_{\mu\nu}^{(0)} F^{(0)\mu\nu} 
+ J^\mu A_\mu^{(0)} \nonumber \\
&& + \sum_{j=1}^{N_f} \sum_{n=-\infty}^{\infty} \mathcal{L}_{\psi_j}^{(n)} \left[ {\psi_j^{(n)}} \right] ,
\label{eq:L-4D-tot-1}
\ea
with
\be
\mathcal{L}_{\psi_j}^{(n)}[\psi] =  \bar{\psi} \left[ i \gamma^\mu \partial_\mu - m_j 
- i \zeta_{j,n} \gamma^5 \right] \psi ,
\label{eq:L-psi-j-n}
\ee
and
\be
J^\mu = - \frac{1}{2\pi Rf} \sum_{j=1}^{N_f} \sum_{n=-\infty}^{\infty} q_j \bar\psi_j^{(n)} \gamma^\mu
\psi_j^{(n)} .
\ee
The massless dark photon $A_\mu^{(0)}$ is coupled to the fermions through the current 
$J^\mu$ but it is still free.
The Lagrangian (\ref{eq:L-4D-tot-1}) obeys the 4D gauge invariance
$A_\mu^{(0)} \to A_\mu^{(0)} - 2 \pi Rf \partial_\mu\alpha$, $\psi_j^{(n)} \to e^{i q_j \alpha} \psi_j^{(n)}$.
In the following we shall integrate over the vacuum dark fermions, so that $\partial_\mu J^\mu$ does not
necessarily vanish as the fermions modes are not on-shell.
However, the integration over the longitudinal mode, $\tilde{A}_\mu^{(0)}(k) \propto k_\mu$
in Fourier space, is not physical as it is a gauge mode.
Then, to only integrate over the physical degrees of freedom transverse to $k_\mu$, 
we impose the Lorenz gauge condition $\partial^\mu A_\mu^{(0)}=0$ \cite{Srednicki_2007}.
This provides a well-defined invertible photon propagator $G_{\mu\nu}$, which reads in Fourier
space as
\be
\tilde G_{\mu\nu}(p) = \frac{1}{p^2+i\epsilon} \left[ \eta_{\mu\nu} - \frac{p_\mu p_\nu}{p^2} \right] .
\ee
Then, the Gaussian integration over the transverse modes of $A_{\mu}^{(0)}$ leads to the Lagrangian
\ba
&& \mathcal{L}_{4D} \left[ \psi_j^{(n)},\phi \right] = \frac{1}{2} \partial_\mu \phi \partial^\mu \phi  
 + \sum_{j=1}^{N_f} \sum_{n=-\infty}^{\infty} \mathcal{L}_{\psi_j}^{(n)} \left[ {\psi_j^{(n)}} \right]
 \nonumber \\
&& \hspace{1cm} + \frac{1}{2} \int d^4 x' \, J^\mu(x) G_{\mu\nu}(x-x') J^\nu(x') .
\label{eq:L-4D-tot-2}
\ea
Integrating over the dark fermions, we obtain the dark matter scalar field Lagrangian
\be
\mathcal{L}(\phi) = \frac{1}{2} \partial_\mu \phi \partial^\mu \phi - V(\phi) ,
\label{eq:L-phi-dark-matter}
\ee
where the axion potential $V(\phi)$ is defined by
\be
e^{-i \int d^4x V(\phi)} = \int \prod_{j=1}^{N_f} \prod_{n=-\infty}^{\infty} {\cal D} \bar\psi_j^{(n)}
{\cal D} \psi_j^{(n)} \, e^{i (S_0+S_I)} ,
\label{eq:V-phi-def}
\ee
with
\be
S_0 = \int d^4 x \, \sum_{j=1}^{N_f} \sum_{n=-\infty}^{\infty} 
\mathcal{L}_{\psi_j}^{(n)}[\psi_j^{(n)}] ,
\label{eq:S0}
\ee
\be
S_I = \frac{1}{2} \int d^4 x \, d^4x' \,  J^\mu(x) G_{\mu\nu}(x-x') J^\nu(x') .
\label{eq:SI}
\ee
If we neglect the spatial dependence of $\phi$ in $S_0$
(we will check the validity of this approximation in section~\ref{sec:derivative}
below), by expressing each Dirac spinor $\psi$ in terms of the Weyl left- and right-handed spinors
$\psi_L$ and $\psi_R$ and writing $m+i \zeta = M e^{i\theta}$, we can check that the phase 
$\theta$ can be absorbed by the phase shift $\psi_R \to e^{-i\theta} \psi_R$.
The quartic action $S_I$ is also invariant with respect to this transformation.
Therefore, with this phase redefinition, we recover the standard Dirac action for $S_0$, with
\be
\mathcal{L}_{\psi_j}^{(n)}[\psi] =  i \bar{\psi} \gamma^\mu \partial_\mu \psi - M_{j,n} \bar{\psi} \psi ,
\label{eq:L-psi-j-n-M}
\ee
and
\be
M_{j,n} =\sqrt{ m_j^2 + \zeta_{j,n}^2} ,
\label{eq:Mjn}
\ee
while $S_I$ is not changed.
We treat the quartic term $S_I$ as a small correction in the path integral (\ref{eq:V-phi-def})
and we expand $e^{iS_I}$ to first order. This gives \cite{Schwartz_2013}
\be
e^{-i \int d^4x V(\phi)} = ( 1 + i \langle S_I \rangle ) \prod_{j,n} 
\det ( i \gamma^\mu \partial_\mu - M_{j,n}) ,
\label{eq:V-eff-def}
\ee
which we can write as
\be
V(\phi) = V_0(\phi) + V_I(\phi) ,
\ee
with
\be
V_0(\phi) =  \sum_{j,n} \, i \, {\rm Tr} \left[ \langle x | \ln \left( i \gamma^\mu \partial_\mu - M_{j,n} 
\right) | x \rangle \right] ,
\label{eq:L-4D-Tr-ln}
\ee
and at linear order over $S_I$,
\ba
&& V_I(\phi) = - \frac{1}{2 (2\pi R f)^2} \int \frac{d^4p d^4 k_1 d^4 k_2}{(2\pi)^8} \tilde G_{\mu\nu}(p)
\bigg[ \delta_D(p) \nonumber \\
&& \times \sum_{j,n} \sum_{j',n'} q_j q_{j'} {\rm Tr}[\gamma^\mu \tilde S_{j,n}(k_1) ] 
{\rm Tr}[\gamma^\nu \tilde S_{j',n'}(k_2) ] \nonumber \\
&& - \delta_D(p \! - \! k_1 \! + \! k_2) \sum_{j,n} q_j^2
{\rm Tr}[\gamma^\mu \tilde S_{j,n}(k_1) \gamma^\nu \tilde S_{j,n}(k_2) \bigg] , \;\;\;
\ea
where $S_{j,n}$ is the Feynman propagator associated with the fermion $\psi_j^{(n)}$ of
mass $M_{j,n}$ and we wrote the integrals in Fourier space.
Using properties of the traces of $\gamma$ matrices and the expression of the fermion
propagator, this reads as
\ba
&& V_I(\phi) =  \frac{2}{(2\pi R f)^2} \int \frac{d^4p d^4 k_1 d^4 k_2}{(2\pi)^8} \tilde G_{\mu\nu}(p)
\bigg[  \sum_{j,n} \sum_{j',n'} \nonumber \\
&& \frac{4 \delta_D(p) q_j q_{j'} k_1^\mu k_2^\nu}{(k_1^2-M_{j,n}^2+i\epsilon) 
(k_2^2-M_{j',n'}^2+i\epsilon)} - \delta_D(p \! - \! k_1 \! - \! k_2)  \nonumber \\
&& \times \sum_{j,n} q_j^2 \frac{ (M_{j,n}^2 + k_1 \cdot k_2) \eta^{\mu\nu} 
- 2 k_1^\mu k_2^\nu}{(k_1^2-M_{j,n}^2+i\epsilon) (k_2^2-M_{j,n}^2+i\epsilon)} \bigg] . \;\;\;
\label{eq:V-I-1}
\ea
The contribution from the first term in the bracket vanishes by parity over $k_1$ or $k_2$.
Using the behavior $\tilde G_{\mu\nu}(p) \sim 1/p^2$, the second term gives a quartic UV divergence
with the cutoff $\Lambda$,
\be
V_I \sim \frac{\Lambda^4}{R^2f^2} , \;\;\; 
\frac{\partial V_I}{\partial M_{j,n}} \sim \frac{\Lambda^2 M_{j,n}}{R^2 f^2}  , \;\;\; 
\frac{\partial^2 V_I}{\partial M_{j,n}^2} \sim \frac{\Lambda^2+M_{j,n}^2}{R^2 f^2} .
\label{eq:V-I-Lambda}
\ee
As seen in Section~\ref{sec:cutoff-Lambda} below,
this will be negligible as compared with $V_0$ even for a cutoff $\Lambda$ that is greater than the
Hubble rate $H_I$ during the inflationary era.
Therefore, in the following we neglect $V_I$ and we focus on the potential $V_0$. 
Thus, each Kaluza-Klein tower associated with a 5D fermion $\psi_j$ generates 
a potential for the pseudo-scalar $\phi$, through the Hosotani mechanism
\cite{Hosotani:1983xw} in the one-loop term $V_0$.
Each contribution is periodic over $\phi$, with the period $2\pi f/q_j$, as $\phi \to \phi+2\pi f/q_j$
is absorbed by the change of index $n \to n-1$ in the infinite Kaluza-Klein sum,
as seen in the expression (\ref{eq:zeta-def}) for $\zeta_{j,n}$.

Thus, the dark matter potential $V(\phi)$ only arises from the fermion fluctuations
in this model. We do not need to introduce any ad-hoc potential in the action
(\ref{eq:L-5D}) or to rely on non-perturbative instanton computations. 
This provides a very simple and explicit setup to generate a dark matter model,
which allows us in the following sections to check that it satisfies theoretical
and observational constraints from the inflationary era to current times.

\subsection{Dark matter potential}
\label{sec:dark-matter-potential}

As seen in appendix~\ref{app:KK-modes}, using standard techniques of quantum field theory
\cite{Schwartz_2013}, the potential (\ref{eq:L-4D-Tr-ln}) can be computed for a constant $\phi$ and
from Eq.(\ref{eq:L-a-m-2}) we obtain
\be
V(\phi) = \sum_{j=1}^{N_f} \sum_{n=-\infty}^{\infty} \frac{1}{8\pi^2}  \int_0^{\infty} \frac{ds}{s^3} \, 
e^{- s M_{j,n}^2 } .
\label{eq:V-a}
\ee
Here and in the following we omit the subscript ``$0$'' in $V_0$.
This expression displays ultraviolet (UV) divergences corresponding to poles at $s=0$. 
A formal expansion of the exponential leads to
\be 
V(\phi) = \sum_j \sum_{n=-\infty}^{\infty} \frac{1}{8\pi^2}  \int_0^{\infty} \frac{ds}{s^3} \, 
\sum_{p=0}^\infty  \frac{(-s M_{j,n}^2)^p}{p!} .
\ee
Commuting the sum over $p$ and the integral over $s$, the first three terms 
$p=0,1,2$ would give UV divergences, whereas the terms $p \geq 2$ would give infrared (IR)
divergences. The latter are merely an artefact of the expansion as the exponential in (\ref{eq:V-a}),
but the UV divergences require some care and a regularization method. 
These expressions could be regularized for instance by introducing 
massive Pauli-Villars fields whose role would be to cancel the UV divergences \cite{Srednicki_2007}. 
This would require fermionic and bosonic fields whose mass differences would cancel 
the three UV divergences associated with $p \leq 2$.
A simpler procedure closer to usual treatments in 4D consists in cutting the $s$ integral as
\be
V_\epsilon(\phi) = \sum_j \sum_{n=-\infty}^{\infty} \frac{1}{8\pi^2} 
\int_\epsilon^{\infty} \frac{ds}{s^3} \, e^{- s M_{j,n}^2 } .
\label{Vreg}
\ee
Notwithstanding the extraneous IR divergences for $p\ge 3$, the formal UV divergences now become
\ba
&& \frac{1}{8\pi^2}  \int_\epsilon ^{\infty} {ds} \, 
\left (\frac{1}{s^3 } - \frac{ M_{j,n}^2}{ s^2 } + \frac{M_{j,n}^4}{2s} \right)\nonumber \\ 
&&= \frac{1}{8\pi^2} \left (\frac{1}{2\epsilon^2} - \frac{ M_{j,n}^2}{ \epsilon} 
+ \frac{M_{j,n}^4}{2}\ln \frac{s_0}{\epsilon}\right) .
\label{eq:ln-UV}
\ea
One can recognize the usual quartic, quadratic and logarithmic divergences of the vacuum energy
where $\epsilon=\Lambda^{-2}$ plays the role of a UV cutoff and we introduced 
$s_0=\Lambda_0^{-2}$ in the logarithmic case, i.e.
a spurious IR cutoff $s_0$ to make the integral finite. 
This reads\footnote{In the dimensional regularization scheme the quartic and quadratic
divergences disappear leaving only the Coleman-Weinberg logarithmic contribution. 
The factor of $1/16\pi^2$ can be retrieved from the standard formulae when noticing
that there are two degrees of freedom for each particle and antiparticle.}
\be
\delta V =  \sum_j \sum_{n=-\infty}^{\infty}  
\left (\frac{\Lambda^4}{16\pi^2} - \frac{ \Lambda^2 M_{j,n}^2}{8\pi^2} 
+ \frac{M_{j,n}^4}{16\pi^2}\ln \frac{\Lambda^2}{\Lambda^2_0}\right) .
\ee
Of course, these expressions are only formal and infinite due to the sum over 
the Kaluza-Klein modes. In fact, the sum over the Kaluza-Klein is harmless if one avoids
expanding the exponential (\ref{Vreg}).
Commuting the $s$ integral and the Kaluza-Klein sum in this finite expression and
using the Poisson summation formula we obtain
\ba
V(\phi) & = & \sum_j \frac{R}{8\pi^{3/2}} \sum_{\ell=-\infty}^{\infty} e^{-i 2\pi \ell (\alpha_j+q_j\phi/(2\pi f)) }
\nonumber \\
&& \times \int_\epsilon^{\infty} ds \; s^{-7/2} e^{-s m_j^2 - (\ell \pi R)^2/s} .
\label{eq:V-s-7/2}
\ea
The term $\ell = 0$ diverges for $\epsilon \to 0$ but it does not depend on $\phi$.
This is a crucial step in the calculation, which allows us to renormalize the potential
in a simple fashion by removing this field-independent constant. 
We can now safely remove the UV cutoff $\epsilon\to 0$ and the integration over $s$
gives the finite result
\ba
V(\phi) & = & \sum_j \frac{1}{32 \pi^6 R^4} \sum_{\substack{\ell = - \infty \\ \ell\neq 0}}^{\infty} 
e^{- | \ell | 2\pi m_j R - i \ell (q_j \phi/f + 2\pi \alpha_j ) } \nonumber \\
&& \times | \ell |^{-5} [ 3 + 6 \pi m_j R | \ell | + (2 \pi m_j R \ell )^2 ] ,
\ea
where the sum over the Kaluza-Klein modes has been properly dealt with.
This also reads as
\ba
V(\phi) & = & \sum_{j=1}^{N_f} \frac{1}{16 \pi^6 R^4} \sum_{\ell = 1}^{\infty} 
e^{- \ell 2\pi m_j R} \cos[\ell (q_j \phi/f + 2\pi \alpha_j ) ] \nonumber \\
&& \times \left( \frac{3}{\ell^5} + \frac{6 \pi m_j R}{\ell^4} 
+ \frac{(2 \pi m_j R)^2}{\ell^3} \right) ,
\label{eq:Va-cos}
\ea
or as
\ba
V(\phi) & = & \sum_{j=1}^{N_f} \frac{1}{16 \pi^6 R^4} {\rm Re} \left[ 3 \, {\rm Li}_5(z_j) 
+ 6 \pi m_j R \, {\rm Li}_4(z_j) \right. \nonumber \\
&& \left. + (2 \pi m_j R)^2 {\rm Li}_3(z_j) \right] , 
\label{eq:Va-Li}
\ea
where ${\rm Li}_s(z)$ are the polylogarithms \cite{olver2010nist}
and we introduced
\be
z_j = e^{-2\pi m_j R + i ( q_j \phi/f + 2\pi \alpha_j ) } .
\ee
The periodicity of each fermion contribution to the potential $V(\phi)$ is a remnant of the
5D gauge invariance of the Lagrangian (\ref{eq:L-5D}), combined with the compactification
on $S^1$.
Except for the Scherk-Schwarz twists $\alpha_j$, this potential also appeared in
\cite{Delgado1999}, in the context of supersymmetric models, and in \cite{Fan2016},
for models of ultralight dark matter with repulsive self-interactions.

The potential (\ref{eq:Va-Li}) is the lowest order in a derivative expansion of 
Eq.(\ref{eq:L-4D-Tr-ln}).
We computed the logarithmic trace neglecting the spacetime dependence of
$\phi(x)$ and substituted into the result (\ref{eq:Va-Li}) the local value $\phi(x)$,
to obtain the potential of our dark matter pseudoscalar field $\phi$.
We will consider in section~\ref{sec:derivative} the first corrections due to gradients of $\phi$.
As expected, we will find in Eq.(\ref{eq:derivative-condition}) that for sufficiently large dark fermion
masses $m_j$ the approximation (\ref{eq:Va-Li}) holds at all cosmological times and scales after the field
$\phi$ has started oscillating in its potential valley and behaving like dark matter.
This is because large masses $m_j$ lead to an exponential cutoff beyond scales of the order of
$1/m_j$ in the fermion-mediated interactions of the scalar field $\phi$.
This makes the potential $V(\phi)$ local in the 4D spacetime.

\section{Building repulsive self-interactions}
\label{sec:building-repulsive}

\subsection{One fermion}

The potential (\ref{eq:Va-cos}) is dominated by the first term $\ell =1$, as higher orders
are suppressed by the exponential damping term and the inverse powers of $\ell$.
Therefore, for a single fermion we obtain at leading order a simple cosine potential,
\be
V(\phi) \simeq \frac{1}{16 \pi^6 R^4} g(m R)  \cos( q \phi/f + 2\pi\alpha) ,
\label{eq:Va-1-fermion}
\ee
where we introduced
\be
g(x) = e^{- 2\pi x} [ 3 + 6 \pi x + (2 \pi x)^2 ] ,
\label{eq:g-def}
\ee
which is a decreasing function of $x$.
As for usual axion potentials \cite{Kim:1986ax,Marsh:2015xka}, 
which are also dominated by a simple cosine term, expanding around a minimum gives
a negative quartic term.
This is why most potentials for ultralight dark matter derived from string or particle physics
models display attractive self-interactions \cite{Fan2016}.
We can check that this remains true for the exact potential (\ref{eq:Va-Li}), where we resummed
all orders in $\ell$.
Therefore, models with only one fermion always lead to attractive self-interactions.
Repulsive self-interactions will be achievable as soon as there are two fermions in the system.

\subsection{Two fermions}
\label{sec:2-fermions}

In the case of two fermions it is easy to design repulsive self-interactions,
using the balance between the two leading cosines.
Keeping only the $\ell=1$ term as the terms for $\ell \ge 2$ are exponentially suppressed for 
$m_j R\gtrsim 1$, we have
\ba
V(\phi) & \simeq & \frac{1}{16 \pi^6 R^4} \left[ g(m_1 R) \cos(q_1 \phi/f + 2 \pi \alpha_1) 
\right. \nonumber\\
&& \left. + g(m_2 R) \cos(q_2 \phi/f + 2 \pi \alpha_2) \right] .
\label{eq:Va-2-fermions}
\ea
Let us consider $m_1 < m_2$, with $g(m_1 R) > g(m_2 R)$.
This means that the potential is dominated by the first fermion and choosing $\alpha_1 = 1/2$
the first cosine has a minimum at $\phi=0$, with a negative quartic term.
We can choose $q_1=1$ without loss of generality, as this amounts to a rescaling of $g_5$.
Then, for $q_2 > 1$ higher-order derivatives of the potential are increasingly
dominated by the second cosine, as each derivative with respect to $\phi$ gives rise to a
multiplicative factor $q_2$.
Choosing $\alpha_2=0$, so that the quartic term of the second cosine is positive, there is
a range of $q_2$ such that the second derivative of the potential is still dominated by the
first cosine but the fourth derivative is dominated by the second cosine.
This provides a model with $V''(\phi)>0$ and $V^{(4)}(\phi)>0$ at $\phi=0$.

Thus, let us consider more precisely the case
\be
\alpha_1 = 1/2 , \;\; \alpha_2 = 0  , \;\; q_1=1 .
\label{eq:alpha1-alpha2}
\ee
This corresponds to anti-periodic and periodic boundary conditions for the 5D fermions $\psi_1$ and
$\psi_2$,
\ba
&& \psi_1(x,y+2\pi R) = - \psi_1(x,y+2\pi R)  , \nonumber \\ 
&& \psi_2(x,y+2\pi R) = \psi_2(x,y+2\pi R) .
\label{eq:anti-periodic}
\ea
Going back to the resummed expression (\ref{eq:Va-Li}), where we take into account
all orders in $\ell$, the potential reads as
\be
V(\phi) = \frac{1}{16 \pi^6 R^4} \left[ h_5(m_1 R, \phi/f+\pi) + h_5(m_2 R, q_2 \phi/f) \right] , 
\label{eq:Va-hn}
\ee
where we introduced the functions
\ba
h_n(x,\theta) & = & {\rm Re} \left[ 3 {\rm Li}_n(e^{-2\pi x+i\theta}) + 6 \pi x  
{\rm Li}_{n-1}(e^{-2\pi x+i\theta}) \right.
\nonumber \\
&& \left. + (2 \pi x)^2  {\rm Li}_{n-2}(e^{-2\pi x+i\theta}) \right] .
\ea
This potential is even, $V(-\phi)=V(\phi)$.
Defining the mass and quartic coupling of the 4D axion as\footnote{The Coleman-Weinberg 
correction to this potential due to the quantum fluctuations of $\phi$ are negligible as soon as 
$f\gg R^{-1}$, which is explicitly verified in our model, see Eq.(\ref{eq:fR}).}
\be
V(\phi) = V_0 + \frac{1}{2} m^2 \phi^2 + \frac{1}{4} \lambda_4 \phi^4 + \dots , 
\label{eq:V-phi-Taylor}
\ee
this gives
\ba
&& m^2 = \frac{h_3^-(m_1 R) - q_2^2 h_3^+(m_2 R)}{16 \pi^6 R^4 f^2} , \nonumber \\
&& \lambda_4 = \frac{q_2^4 h_1^+(m_2 R) - h_1^-(m_1 R)}{96 \pi^6 R^4 f^4} ,
\ea
with
\be
h_n^+(x) = h_n(x,0) , \;\; h_n^-(x) = - h_n(x,\pi) .
\ee 
Repulsive self-interactions correspond to the range
\be
m^2>0, \; \lambda_4 > 0 : \;\; \sqrt{\frac{h_1^-(m_1 R)}{h_1^+(m_2 R)}} < q_2^2 
< \frac{h_3^-(m_1 R)}{h_3^+(m_2R)} .
\label{eq:q2-range}
\ee
For large fermion masses, this range behaves at leading order as
\be
m_i R \gg 1 : \;\; e^{\pi R (m_2-m_1)} \lesssim q_2^2 \lesssim e^{2\pi R (m_2-m_1)} ,
\label{eq:range-q2-approx}
\ee
and it exists as soon as $m_2 > m_1$.

In the case $m_1=m_2=0$, only the term ${\rm Li}_5(z)$ remains in Eq.(\ref{eq:Va-Li})
and we obtain
\be
V(\phi) = \frac{1}{16 \pi^6 R^4} \left[ \bar h_5(\phi/f+\pi) + \bar h_5(q_2 \phi/f) \right] , 
\label{eq:Va-m1=0}
\ee
with
\be
\bar h_n(\theta) = {\rm Re} \left[ 3 {\rm Li}_n(e^{i\theta}) \right] .
\ee
This gives
\be
m_1=m_2=0: \;\;\; m^2 = \frac{3 \zeta(3) (3-4 q_2^2)}{64 \pi^6 R^4 f^2} , \;\; \lambda_4=\infty ,
\label{eq:m1-m2-0}
\ee
where $\zeta(x)$ is the Riemann zeta function.
The potential has a singularity at $\phi=0$, $V(\phi) \supset - (\phi/f)^4 \ln(\phi/f)$, which leads to
an infinite fourth derivative.
This gives a repulsive potential for $|q_2| < \sqrt{3}/2$,
with an effective coupling that exhibits a logarithmic dependence on the dark matter density.
We shall consider in more details this case in Section~\ref{sec:low-mass} below.

\begin{figure}
\centering
\includegraphics[height=6.5cm,width=0.45\textwidth]{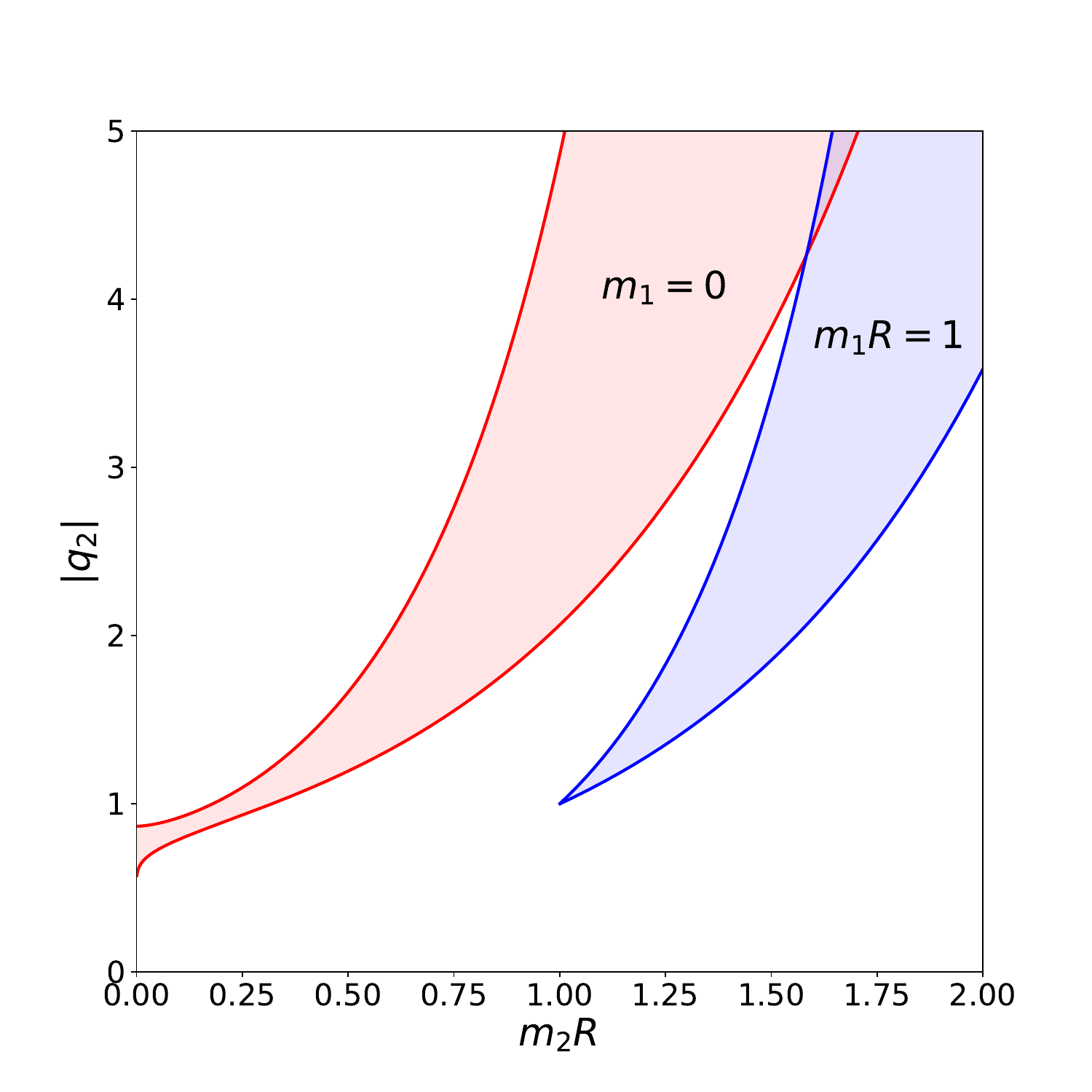}
\caption{
The shaded regions are the domains in the plane $(m_2R,|q_2|)$ where the
quartic self-interactions are repulsive, for the cases $m_1=0$ (red) and $m_1 R=1$ (blue).
}
\label{fig:q2}
\end{figure}

\begin{figure}
\centering
\includegraphics[height=6.5cm,width=0.45\textwidth]{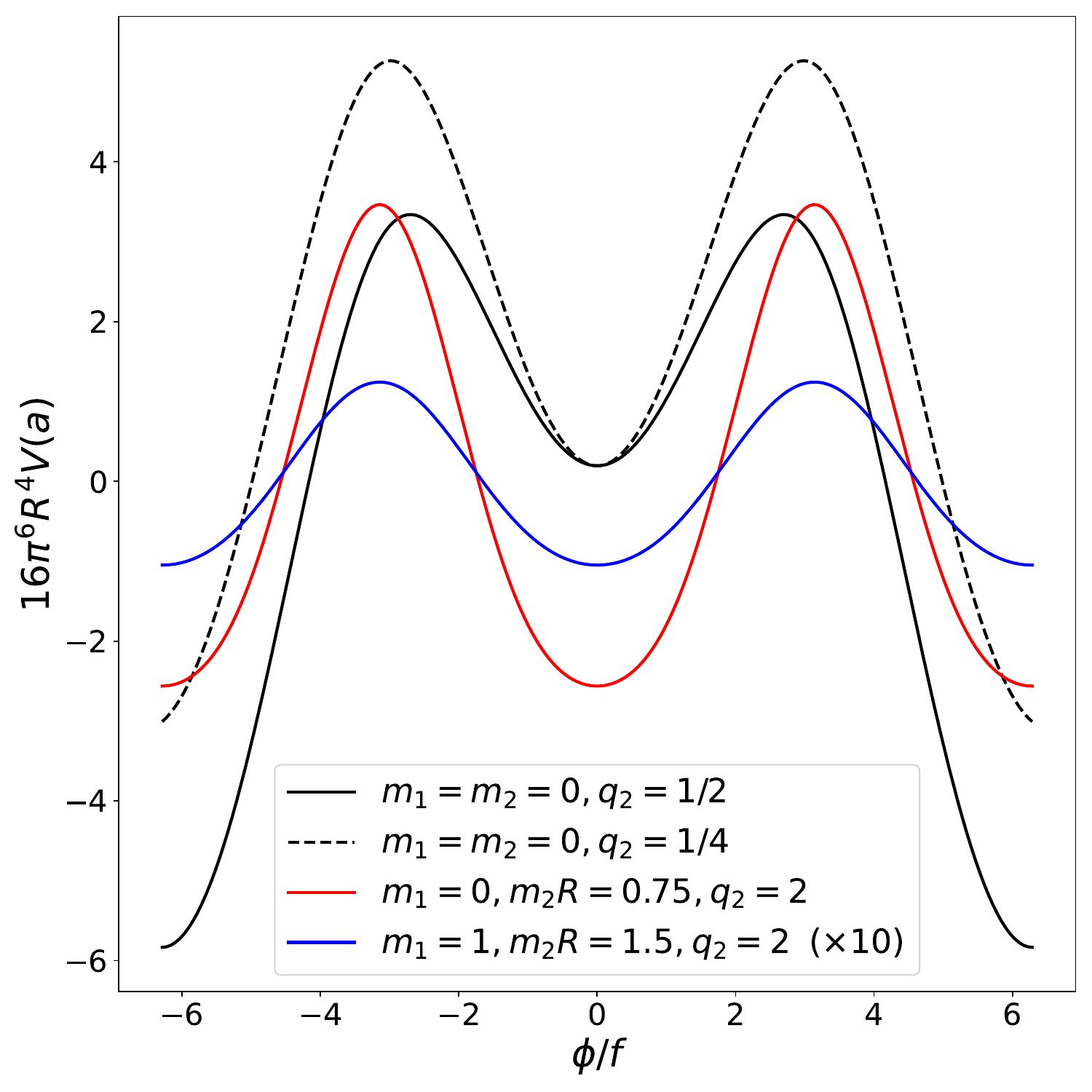}
\caption{
Potential $V(\phi)$, normalized to $1/(16 \pi^6 R^4)$, for the cases
$\{m_1=m_2=0\}$ (solid and dashed black curves), $\{m_1=0,m_2 R=0.75\}$ (red curve) 
and $\{m_1R=1, m_2R=1.5\}$ (blue curve), for values of $q_2$ as indicated in the figure.
For the case $\{m_1R=1, m_2R=1.5\}$ the potential is multiplied by a factor $10$ for clarity.
}
\label{fig:Va}
\end{figure}

We show in Fig.~\ref{fig:q2} the domains in the plane $(m_2R,|q_2|)$ where $m^2>0$ and 
$\lambda_4>0$, for the cases $m_1=0$ and $m_1 R=1$.
For the case $m_1R=1$, we recover the behavior (\ref{eq:range-q2-approx}), with a domain
the extends over $m_2>m_1$ and a charge $|q_2|$ that follows an exponential growth with
the mass difference $(m_2-m_1)$.
In the case $m_1=0$, we recover the same behavior (\ref{eq:range-q2-approx}) for $m_2 R \gtrsim 1$.
For low fermion masses, it is necessary to go beyond the leading order (\ref{eq:range-q2-approx})
and to work with the resummed potential (\ref{eq:Va-hn}).
In agreement with (\ref{eq:m1-m2-0}), this leads to a finite-size domain at low fermion masses,
down to the case $m_1=m_2=0$.
Thus, the 5D shift-symmetric model with vanishing fermion masses still provides repulsive
self-interactions if $|q_2|<\sqrt{3}/2$.

We show in Fig.~\ref{fig:Va} the potential $V(\phi)$ for the three cases
$\{m_1=m_2=0\}$, $\{m_1=0,m_2 R=0.75\}$ and $\{m_1R=1, m_2R=1.5\}$, with values of $q_2$
that correspond to repulsive quartic self-interactions.
The two terms in the potential (\ref{eq:Va-hn}), associated with the two fermions,
have periodicities in $\phi/f$ of $2\pi$ and $2\pi/|q_2|$.
This only gives a periodic potential if $q_2$ is rational.
In the cases $\{m_1=0,m_2 R=0.75\}$ and $\{m_1R=1, m_2R=1.5\}$ shown in the figure,
with $q_2=2$, the potential is periodic with period $P=2\pi$ over $\phi/f$.
In the case $m_1=m_2=0$, where repulsive self-interactions require $|q_2|<\sqrt{3}/2$,
the period is greater than $2\pi$.
For the cases $q_2=1/2$ and $1/4$ shown in the figure we have $P=4\pi$ and $8\pi$.
Moreover, in this case the minimum at $\phi=0$ is only a local minimum, while the
global minimum at $\phi=\pi/q_2$ (for even integer $q_2$) has attractive self-interactions.
We shall study the lifetime of this metastable state in Section~\ref{sec:low-mass} below.

The potential (\ref{eq:Va-cos}) was already used as a tool to build ultralight dark matter
models with repulsive self-interactions in \cite{Fan2016}, without the Scherk-Schwarz twists 
$\alpha_j$ and focusing on the regime $m_j R \ll 1$.
It was found that repulsive self-interactions only appear in small islands of the parameter
space, with at least two bosons and two fermions.
Our model (\ref{eq:Va-hn}) is in this sense more economical, as we only require two fermions.
This is because the Scherk-Schwarz twists $\alpha_j$ essentially allows us to change the 
relative sign of the two leading cosines (instead of combining one fermion with one boson)
and we permit different masses $m_j$. Then, for given $(m_1,q_1)$ the domain associated
with repulsive self-interactions shows as extended bands in the space $(m_2,q_2)$.
This suggests that ultralight dark matter models are not so far-fetched as could have been assumed.
We also extend on \cite{Fan2016} by paying attention to several additional theoretical and 
observational constraints on these models, such as isocurvature bounds, the cosmological gravitational
particle production of dark fermions and photons during inflation, and the corrections to the
potential (\ref{eq:Va-Li}) in a derivative expansion due to gradients of the dark matter field.

\section{Late-time constraints on the parameters}
\label{sec:late-time-constraints}

\subsection{Dark matter mass and coupling constant}

In the following we will consider models in the class (\ref{eq:alpha1-alpha2}),
with two fermions at a single 5D fermion mass scale $m_\psi$
and charges of order unity,
\be
m_1 \sim m_2 \sim m_\psi , \;\;\; q_1 \sim q_2 \sim 1 .
\label{eq:m1-m2-mpsi}
\ee
This corresponds to repulsive self-interactions in the range (\ref{eq:q2-range}), with
$(m_2-m_1) R \lesssim 1$.
As can be seen from Eq.(\ref{eq:Va-cos}),
this gives a dark matter mass $m$ and coupling $\lambda_4$
\be
m^2 \sim \frac{3+(2\pi R m_\psi)^2}{16 \pi^6 R^4 f^2} e^{-2\pi R m_\psi} , \;\;\;
\lambda_4 \sim \frac{m^2}{6 f^2} .
\label{eq:m2-lambda4}
\ee
In these expressions, one can see that several dimensionless quantities matter. The axion mass is suppressed as soon as $m_\psi R \gtrsim 1$, although this product cannot be too large
so that $fR \gg 1$ in Eq.(\ref{eq:fR}) below.
The mass is also reduced by this large factor $fR$. 
Finally a small self-interaction coupling is obtained as $m\ll f$, see Eq.(\ref{eq:m-over-f})
below.

\subsection{Misalignment mechanism and dark matter abundance}

For small values of $\phi$, the potential $V(\phi)$ becomes quadratic with a small quartic 
correction. At late times, when the scalar mass $m$ is greater than the Hubble rate $H$,
the scalar field oscillates in an almost harmonic potential. Averaging over the oscillations
this leads to a vanishing effective pressure and the background density decreases in the matter
era as $\rho \propto a^{-3}$, which means that it behaves like dark matter \cite{Brax:2019fzb}.
The small repulsive quartic self-interaction leads to a small effective pressure
that only plays a role on small scales, where it can suppress the matter power spectrum
as compared with CDM and lead to the formation of stable hydrostatic equilibria, known
as solitons or boson stars, on galactic or astrophysical scales \cite{Goodman:2000tg,Chavanis:2011zi}.
The scalar field $\phi$ starts oscillating at time $t_{\rm osc}$, when the Hubble expansion rate
is approximately $H_{\rm osc} \simeq m/3$. Defining the initial value $\phi_i = \theta_i f$,
with $|\theta_i| \lesssim \pi$, the usual misalignment mechanism 
\cite{Marsh:2015xka,OHare:2024} gives the dark matter density today
\be
\Omega_\phi = \frac{g_{\star s}(T_0)}{6 g_{\star s}(T_{\rm osc})} 
\left( \frac{\pi^2 g_{\star}(T_{\rm osc})}{10} \right)^{3/4} \frac{\theta_i^2 f^2 m^{1/2} T_0^3}
{M_{\rm Pl}^{7/2} H_0^2} ,
\label{eq:Omega-phi-f-m}
\ee
which also reads as
\be
\Omega_\phi \simeq 0.37 \; \theta_i^2 \left( \frac{f}{10^{13} \, {\rm GeV}} \right)^2
\left( \frac{m}{10^{-6} \; {\rm eV}} \right)^{1/2} .
\ee
Thus, the field $\phi$ makes up all of the dark matter if
\be
f = 10^{13} \, {\rm GeV} \; \theta_i^{-1} \left( \frac{m}{10^{-6} \, {\rm eV}} \right)^{-1/4} .
\label{eq:f-abundance}
\ee
Using Eq.(\ref{eq:m2-lambda4}), this gives
\ba
m & \simeq & 3.5 \times 10^{28} \; {\rm eV} \; \theta_i^2 ( R \,\times 10^{13} \, {\rm GeV} )^{-8/3}
\nonumber \\
&& \times [ 3 + (2\pi R m_\psi)^2 ]^{2/3} e^{-4 \pi R m_\psi/3} 
\label{eq:m-m_psi}
\ea
and
\be
\lambda_4 \simeq 2 \times 10^{-57} \theta_i^2 \left( \frac{m}{10^{-6} \, {\rm eV}} \right)^{5/2} 
\ll 1.
\label{eq:lambda4-m}
\ee
The ratio $m/f$ is then given by
\be
\frac{m}{f} \simeq 10^{-28} \, \theta_i \left( \frac{m}{10^{-6} \, {\rm eV}} \right)^{5/4}  \ll 1 ,
\label{eq:m-over-f} 
\ee
which shows that $m \ll f$.

The relation (\ref{eq:m-m_psi}) also provides the scale $R$ of the fifth dimension in terms
of the dark matter mass $m$ as
\ba
\frac{1}{R} & \sim & 1 \, {\rm GeV} \left( \frac{m}{10^{-6} \, {\rm eV}} \right)^{3/8}
\theta_i^{-3/4} [ 3 + (2\pi R m_\psi)^2 ]^{-1/4} \nonumber \\
&& \times e^{\pi R m_\psi/2} .
\label{eq:R-m}
\ea
This gives for the product $fR$
\ba
fR & \sim & 10^{13} \left( \frac{m}{10^{-6} \, {\rm eV}} \right)^{-5/8} \theta_i^{-1/4}
[ 3 + (2\pi R m_\psi)^2 ]^{1/4} \nonumber \\
&& \times e^{-\pi R m_\psi/2} .
\label{eq:fR}
\ea
Thus, we verify that $f R \gg 1$ if $R \, m_\psi$ is not too large,
\be
f R \gg 1 \;\;\; \mbox{for} \;\;\; R \, m_\psi \lesssim 20 .
\label{eq:fR-Rm_psi}
\ee

The field $\phi$ must start oscillating and behaving like dark matter before radiation-matter equality.
This gives the usual lower bound 
\be
m > 5 \times 10^{-28} \, {\rm eV} .
\ee
We also assumed that the oscillations start in the radiation era, after the end of inflation, which
holds for almost all inflationary scenarios as we focus on ultralight dark matter models
with $m < 1 \, {\rm eV}$.
The Hubble expansion rate $H_I$ at inflation can span the wide range 
$10^{-15} \, {\rm GeV} < H_I < 10^{13} \, {\rm GeV}$, where the lower bound comes from the requirement
that inflation ends before the electroweak transition
while the upper bound is given by the Planck upper bound on B-mode polarisation \cite{Planck2020}.

\subsection{Jeans length and soliton radius}

In the nonrelativistic regime, with $| \phi | \ll f$, when the field is oscillating near its minimum,
the nonlinear Klein-Gordon equation of motion obeyed by the scalar field $\phi$ can be simplified
to a nonlinear Schr\"odinger equation by introducing the complex scalar field $\psi$ as
\cite{Hui:2016ltb,Brax:2019fzb}
\be
\phi(\vec x,t) = \frac{1}{\sqrt{2m}} \left( \psi(\vec x,t) \, e^{-imt} 
+ \psi^\star(\vec x,t) \, e^{imt} \right) .
\ee
This factors out the fast oscillations at angular frequency $m$ from the slower evolution
on astrophysical or cosmological time scales described by $\psi(\vec x,t)$.
Substituting into the Klein-Gordon equation gives in the nonrelativistic regime
a nonlinear Schr\"odinger equation, also known as a Gross-Pitaevskii equation,
\be
i \frac{\partial\psi}{\partial t} = - \frac{\Delta\psi}{2m} + m (\Phi_N + \Phi_I) \psi ,
\label{eq:Schrodinger}
\ee
where we neglected the cosmological expansion, as we focus on scales much smaller than the horizon.
The Newtonian gravitational potential $\Phi_N$ is determined by the Poisson equation, 
$\Delta \Phi_N = 4 \pi {\cal G} (\rho+\rho_b)$, where we included the baryonic density $\rho_b$,
and the dark matter density is $\rho=m|\psi|^2$.
The potential $\Phi_I$ arises from the quartic term in the dark matter potential
(\ref{eq:V-phi-Taylor}), where we neglect higher-order terms that are even smaller for
$| \phi | \ll f$. It is given by \cite{Brax:2019fzb}
\be
\Phi_I(\rho) = \frac{\rho}{\rho_a}  ,  \;\;\; \rho_a = \frac{4 m^4}{3 \lambda_4} 
\sim 8 m^2 f^2 ,
\label{eq:PhI-rho_a}
\ee
where in the last estimate we used Eq.(\ref{eq:m2-lambda4}).
The potential $\Phi_I$ describes the effective pressure due to the repulsive self-interactions.
Using the Madelung transform \cite{Madelung:1927ksh} to write the Schr\"odinger equation
Eq.(\ref{eq:Schrodinger}) in a hydrodynamical form, as a system of continuity and Euler equations, 
one can check that the potential (\ref{eq:PhI-rho_a}) corresponds to an effective pressure
$P_I = \rho^2/(2\rho_a)$, which is a polytropic equation of state of exponent $\gamma=2$.
In particular, this shows that repulsive self-interactions correspond to $\lambda_4>0$,
with $dP_I/d\rho>0$.
Using Eq.(\ref{eq:lambda4-m}), we obtain
\ba
\rho_a & \simeq & ( 0.17 \, {\rm GeV} )^4 \, \theta_i^{-2} \left( \frac{m}{10^{-6} \, {\rm eV}} \right)^{3/2} \nonumber \\
& \simeq & 2.9 \times 10^{36} M_\odot/{\rm pc}^3 \; \theta_i^{-2} 
\left( \frac{m}{10^{-6} \, {\rm eV}} \right)^{3/2} .
\label{eq:rho_a-m}
\ea
The density $\rho_a$ corresponds to the threshold $|\phi| \sim f$ where the quartic term 
becomes of the same order as the quadratic term in the dark matter potential
(\ref{eq:V-phi-Taylor}).
The extremely large value (\ref{eq:rho_a-m}) means that in the late Universe the scalar field
is very close to the minimum of its potential, at $\phi=0$.
Then, the dark matter potential (\ref{eq:V-phi-Taylor}) is dominated by far by the quadratic term
in Eq.(\ref{eq:V-phi-Taylor}) and the quartic term provides a small repulsive pressure that is 
negligible on cosmological scales, as for collisionless dark matter.

However, on small scales the repulsive self-interaction can lead to the formation of 
hydrostatic equilibrium configurations, also called solitons or boson stars (when their size is 
similar to a small compact object),
where the effective pressure $P_I$ balances the dark matter self-gravity.
Neglecting the baryonic component, the equation of hydrostatic equilibrium reads
$\nabla (\Phi_N + \Phi_I) = 0$. This gives the dark matter density profile
\cite{Goodman:2000tg,Chavanis:2011zi,Brax:2019fzb}
\be
\rho(r) = \rho_0 \frac{\sin(\pi r/R_s)}{\pi r/R_s} ,
\label{eq:soliton-rho}
\ee
where the soliton radius $R_s$ is given by
\be
R_s = \sqrt{\frac{\pi}{4 {\cal G} \rho_a}} .
\label{eq:Rsol-def}
\ee
Thus, the radius $R_s$ does not depend on the mass of the soliton but only on the parameters
$m$ and $\lambda_4$ of the dark matter potential.
This also reads as
\be
R_s \simeq 3 \times 10^{-12} \, {\rm pc} \; \theta_i \left( \frac{m}{10^{-6} \, {\rm eV}} \right)^{-3/4} ,
\ee
or
\be
R_s \simeq 8 \times 10^{-2} \, {\rm pc} \; \theta_i \left( \frac{m}{10^{-20} \, {\rm eV}} \right)^{-3/4} .
\label{eq:Rs-pc-2}
\ee
The scale $R_s$ also gives the size of the Jeans length, which does not depend on redshift
in this model \cite{Goodman:2000tg,Peebles_2000,Arbey2003}.

On the other hand, the de Broglie wavelength $\lambda_{\rm dB}$ is given by
\be
\lambda_{\rm dB} = \frac{2\pi}{mv} \sim 5 \; \mbox{pc} \left(\frac{10^{-20}\, {\rm eV}}{m} 
\right) \left( \frac{250 \, {\rm km/s}}{v} \right) ,
\label{eq:lambda-dB}
\ee
where $v$ is the typical virial velocity of the dark matter cloud.
The pressure due to the self-interaction dominates over the so-called quantum pressure
when $R_s>\lambda_{\rm dB}$, that is,
\be
R_s>\lambda_{\rm dB} : \;\; m > 1.5 \times 10^{-13} \, {\rm eV} \, \theta_i^{-4} 
\left( \frac{v}{250 \, {\rm km/s}} \right)^{-4} .
\label{eq:Rs-dB}
\ee
On the other hand, observations of small-scale structures like the Lyman-$\alpha$ forest
provide a lower bound on $\min(R_s,\lambda_{\rm dB})$, which gives the usual lower bound
\cite{Armengaud:2017nkf,Rogers:2020ltq}
\be
m \gtrsim 10^{-21} \, {\rm eV} ,
\label{eq:n-Ly-alpha}
\ee
as the quantum pressure dominates for such low masses.

\subsection{Small-scale tests of gravity}

Gravity has been probed down to scales $r \sim 50 \, \mu m$ \cite{Lee:2020zjt}.
This implies the bound on the radius $R$ of the extra dimension,
\be
R \lesssim 10^{-5} \, {\rm m} , \;\;\; 1/R \geq 0.02 \, {\rm eV} .
\label{eq:gravity-test}
\ee
This is a relatively short distance, especially as compared with astrophysical scales, 
but still a low energy scale as compared to high-energy phenomena such as inflation.

\section{Constraints for a pre-inflationary scenario}
\label{sec:pre-inflationary}

If the 5D theory (\ref{eq:L-5D}) can only be compactified to 4D after inflation,
the scalar field $\phi$ may oscillate around different minima, $\phi_n = n 2\pi f$,
in different Hubble patches.
This will create domain walls.
Even though the dynamics of the universe may lead to a scaling regime, where only
one domain wall remains inside the horizon, such walls have not been observed in the data 
and we will not discuss this possibility in this paper.

Therefore, we focus on models that are already valid during the inflationary era. 
This leads to several requirements, so that the isocurvature fluctuations generated by the
axion-like field $\phi$ do not violate upper bounds derived from CMB data, or 
the dark fermions $\psi_j^{(n)}$ and photons $A_\mu^{(c,n)}$ and $A_\mu^{(s,n)}$
are not gravitationally produced in too large quantities during inflation.
In this Section we check that our model satisfies these constraints.

\subsection{Compactification before inflation}

The 5D theory (\ref{eq:L-5D}) can be compactified to 4D before inflation 
and does not modify the cosmological evolution if
\be
1/R \gg H_I .
\label{eq:R-HI}
\ee
This guarantees that only the zero modes of the 5D theory are excited during inflation and that one can trust
a 4D description at low energy. This is the same condition as applied to string inflation models 
\cite{Baumann:2014nda} where the 10D behavior of string theory
is condensed in a 4D effective action as long as $H_I \ll M_{\rm KK}$ where $M_{\rm KK}$ is the 
compactification scale corresponding to $1/R$ in our 5D model.
If we require inflation to occur before the electroweak phase transition, when
$H_{\rm EW} \sim 10^{-15} \, {\rm GeV}$, we obtain
\be
\frac{1}{R} \gg H_I > H_{\rm EW} \sim 10^{-6} \; {\rm eV} ,
\label{eq:R-EW}
\ee
which is less stringent than the lower bound (\ref{eq:gravity-test}) for very late
inflationary scenarios.

\subsection{Isocurvature bounds}

If the field $\phi$ is physical before inflation, isocurvature quantum fluctuations
due to the light $\phi$ field are generated
during the inflation era \cite{Marsh:2015xka,OHare:2024}, with a power-spectrum amplitude
\be
A_{\rm iso} \simeq \left( \frac{H_I}{\pi f \theta_i} \right)^2 .
\ee
The upper bound from the Planck satellite, $A_{\rm iso} \leq 8 \times 10^{-11}$, gives an upper bound
on the Hubble expansion rate at the time of the inflation,
\be
H_I \leq  3 \times 10^8 \, {\rm GeV} \; \left( \frac{m}{10^{-6} \, {\rm eV}} \right)^{-1/4} .
\label{eq:HI-iso}
\ee
For $m= 1 \, {\rm eV}$ this gives the upper bound $H_I \leq 10^7 {\rm GeV}$.
This is consistent with current data as Planck only gives the upper bound 
$H_I < 6 \times 10^{13} \, {\rm GeV}$ and the scalar-to-tensor ratio has not been measured yet.

\subsection{Constraints from the scale of the fifth dimension}

From Eq.(\ref{eq:R-m}), the gravity bound (\ref{eq:gravity-test}) is satisfied as soon as
\be
m > 3 \times 10^{-23} \; {\rm eV} \; \theta_i^2 ,
\ee
which is obtained by taking $2\pi R m_{\psi} \lesssim 1$.
For larger values of $2\pi R m_{\psi}$ the scale $1/R$ is greater and the lower bound
(\ref{eq:gravity-test}) is even more easily satisfied.
We can see that this condition is always verified for models that pass the Lyman-$\alpha$
constraint (\ref{eq:n-Ly-alpha}).
The inflationary bound (\ref{eq:R-HI}) can be read as an upper bound on the inflation scale,
\be
H_I \ll 1 \, {\rm GeV} \; \left( \frac{m}{10^{-6} \, {\rm eV}} \right)^{3/8}  \theta_i^{-3/4} ,
\ee
in the case $2\pi R m_{\psi} \lesssim 1$. Again, this is only a sufficient condition in the case
$2\pi R m_{\psi} \gg 1$.
This bound is stronger than the isocurvature bound (\ref{eq:HI-iso}) but it is again
fully consistent with current data.

\subsection{Dark fermions production}
\label{sec:Fermion-prod}

\subsubsection{Relic energy density}

After integration over the fifth dimension on the circle of radius $R$, the 4D Lagrangian
(\ref{eq:L-4D-tot}) includes a tower of Kaluza-Klein Dirac fermions $\psi_j^{(n)}$, of mass
(\ref{eq:Mjn}) as in Eq.(\ref{eq:L-psi-j-n-M}).
The background $\psi_j^{(n)}=0$ is a classical solution, as there is no source term in the total
action. However, the cosmological expansion of the Universe gives rise to a time dependence
of the quantum vacuum, which leads to a gravitational production of particles, as for the
production of primordial fluctuations during the inflationary stage \cite{Mukhanov:2005sc}.
This produces a relic energy density that must remain negligible during the cosmological evolution
to be consistent with cosmological probes that measure the abundance of matter and radiation
since big bang nucleosynthesis.
We describe in appendix~\ref{app:production-fermions} the formalism that we use to compute
the energy density of these fermionic degrees of freedom, following \cite{Kolb:2023ydq}.

We work with the conformal time $\eta$ as in Eq.(\ref{eq:FLRW}).
We consider an inflationary stage from initial time $t_i$ to final time $t_e$,
with a de Sitter expansion with constant Hubble expansion rate $H_I$,
\be
t_i < t < t_e : \;\; a(t) = e^{H_I t} = - \frac{1}{H_I \eta} ,
\label{eq:Inflation-a}
\ee
followed by the radiation era until the matter-radiation equality.
We can consider separately each fermion of the Kaluza-Klein tower $\psi_j^{(n)}$.
In curved spacetime the Dirac action becomes Eq.(\ref{eq:S-Psi}). With the rescaling
(\ref{eq:Dirac-eta}) we recover the usual Dirac action except that the mass becomes time dependent.
This gives rise to the change of quantum vacuum and to the particle production,
which vanish for $m_\psi \to 0$ because the action is conformally invariant in the massless case.
As seen in appendix~\ref{app:production-fermions} using the standard Bogolyubov coefficients, 
the energy density at late times reads as \cite{Chung:2011ck,Kolb:2023ydq}
\be
\rho = \frac{4}{a^4} \int \frac{d\vec k}{(2\pi)^3} \omega_k | \tilde\beta_k |^2 
=  \frac{2}{\pi^2 a^4} \int \frac{dk}{k} k^3 \omega_k | \tilde\beta_k |^2 ,
\label{eq:rho-tilde-beta}
\ee
where the frequency $\omega_k$ of the mode of comoving wavenumber $k$ is given by 
Eq.(\ref{eq:omega-k}) and the coefficient $\tilde\beta_k$ is the solution of the system
of equations (\ref{eq:tilde-alpha-beta}), evaluated at late times
$\eta \to \infty$, with the initial condition $(\tilde\alpha_k=1,\tilde\beta_k=0)$
at $\eta \to -\infty$.

In regimes where the particle production is inefficient, we have $\tilde \alpha_k \simeq 1$
and $\tilde\beta_k \simeq 0$ at all times.
Then, at leading order we can integrate the second equation in (\ref{eq:tilde-alpha-beta})
with $\tilde\alpha_k \simeq 1$ as
\be
\tilde\alpha_k \simeq 1 , \;\;\; | \tilde\beta_k | \ll 1 , \;\;\;
\tilde\beta_k \simeq \int d\eta \frac{a'm_\psi k}{2\omega_k^2} e^{-2i\Phi_k} ,
\label{eq:beta-int}
\ee
where $\Phi_k(\eta)$ is given in Eq.(\ref{eq:omega-k}) .
In addition to the initial and final times of the inflation (\ref{eq:Inflation-a}),
we define the other characteristic times $t_m, t_k$ and $t_H$ by
\be
H(t_m) = m_\psi , \;\;\; a(t_k) = \frac{k}{m_\psi} , \;\;\; a(t_H) H(t_H) = k .
\label{eq:tm-tk}
\ee
Before the time $t_k$, the mode is relativistic and we have
\be
t < t_k : \; \omega_k \simeq k, \; \Phi_k \simeq k \int \! \frac{d\ln a}{aH} , \;
\frac{d\tilde\beta_k}{d\ln a} \sim \frac{a m_\psi}{2k} e^{-2i\Phi_k} ,
\ee
while at later times in the nonrelativistic regime we have
\ba
t > t_k : && \;\;\; \omega_k \simeq a m_\psi , \; \Phi_k \simeq m_\psi \int \frac{d\ln a}{H} , 
\nonumber \\
&& \frac{d\tilde\beta_k}{d\ln a} \sim \frac{k}{2 a m_\psi} e^{-2i\Phi_k} .
\ea
We can now describe the creation of particles for both low or high scalar masses, 
$m_\psi < H_I$ or $m_\psi > H_I$.

\subsubsection{Low mass $m_\psi < H_I$}

We first consider the low-mass range $H_{\rm eq} < m_\psi < H_I$, whence $t_e < t_m < t_{\rm eq}$,
where $t_{\rm eq}$ is the time at matter-radiation equality,
with $H_{\rm eq} \simeq 1.6 \times 10^{-37} {\rm GeV}$.
We also consider modes that enter the horizon after the end of inflation, $t_H>t_e$.
Defining the wavenumber $k_m = a(t_m) m_\psi$, we have
\be
k < k_m : \; t_k < t_m < t_H , \;\;\;
k > k_m : \; t_H < t_m < t_k .
\ee
For low wavenumbers $k<k_m$ that became nonrelativistic at a time $t_k<t_m$,
we find that $\tilde\beta_k$ grows and becomes of the order of unity at time $t_k$
when $\Phi_k(t_k) \sim m_\psi/H(t_k) < 1$.
Because the coefficients $\tilde\alpha_k(\eta)$ and $\tilde\beta_k$ introduced in
Eq.(\ref{eq:tilde-alpha-tilde-beta}) satisfy the normalization
$| \tilde\alpha_k |^2 + | \tilde\beta_k |^2 = 1$, we have $| \tilde\beta_k |^2 \leq 1$,
and $| \tilde\beta_k |$ remains of the order of unity afterwards. This gives
\be
a(t_i) m_\psi < k < k_m : \;\; \tilde\beta_k \sim 1 .
\ee
Higher wavenumbers $k>k_m$ enter the horizon at a time $t_H$ with $t_H < t_m < t_k$, 
when $\Phi_k(t_k) \sim 1$ and $\tilde\beta_k \sim a_H m_\psi /(2k)$. 
The growth of the coefficient $\tilde\beta_k$ is damped
at later times by the fast oscillations of the factor $e^{-2i\Phi_k}$ and we obtain
\be
k > k_m : \;\; \tilde\beta_k \sim \left( \frac{k}{k_m} \right)^{-2} ,
\ee
where we used $a_H \propto k^{-1}$ for modes that enter the horizon during the radiation era.
Therefore, we approximate the energy density (\ref{eq:rho-tilde-beta}) today by cutting
the integral at $k_m$, with $|\tilde\beta_k |^2=1/2$ for $k<k_m$ and $|\tilde\beta_k |^2=0$
for $k>k_m$,
\be
\rho(t_0) \simeq \frac{m_\psi k_m^3}{3\pi^2 a_0^3} 
= \frac{m_\psi^4}{3\pi^2} \left( \frac{a(t_m)}{a_0} \right)^3 .
\ee
This gives
\be
\frac{\rho(t_0)}{\rho_{c0}} \simeq 1.6 \times 10^{-21} \left( \frac{m_\psi}{1 \, {\rm GeV}} 
\right)^{5/2} ,
\ee
where $\rho_{c0} \simeq 3.6 \times 10^{-47} \, {\rm Gev}^4$ is the critical density today.
This gives the upper bound on the fermion mass $m_\psi$
\be
\rho(t_0) \ll \rho_{c0} : \;\; m_\psi \ll 2 \times 10^8 \, {\rm GeV} .
\label{eq:mpsi-upper}
\ee
We can check that lower masses, $m_\psi < H_{\rm eq}$, lead to even smaller energy densities,
so that the upper bound (\ref{eq:mpsi-upper}) applies to the full range $m_\psi < H_I$.

\subsubsection{High mass $m_\psi > H_I$}
\label{sec:fermions-high-mass}

If $m_\psi \gg H_I$ the particle production is again inefficient and we can use
Eq.(\ref{eq:beta-int}), which is again dominated by the time $t_k$.
We first consider modes with $t_k < t_e$, which corresponds to $k<k_e$ with
$k_e = a(t_e) m$.
Using Eq.(\ref{eq:Inflation-a}), during the inflationary stage the integral
(\ref{eq:beta-int}) reads
\be
\tilde\beta_k = \frac{1}{2} \int_{-\infty}^0 \frac{du}{1+u^2} e^{2i (m_\psi/H_I) 
[ \sqrt{1+u^2} - {\rm ArcCoth} \sqrt{1+u^2} ]} ,
\label{eq:tilde-beta-u-fermions}
\ee
where we solved for $\Phi_k$ and we made the change of variable $u=H_I k \eta/m_\psi$.
Here we assumed that at the end of inflation $|u| \ll 1$, which corresponds to wavenumbers
$k< a(t_e) m$.
In the limit $m_\psi/H_I \to \infty$, the integral is dominated by the saddle point
$u = - i + \frac{1}{2} e^{i 5 \pi/6} (m_\psi/H_I)^{-2/3}$ and we obtain the exponential
falloff
\be
k < k_e : \;\; \tilde\beta_k \simeq \frac{\pi}{3} e^{-\pi m_\psi/H_I} .
\label{eq:fermion-exp-cutoff}
\ee
Modes at higher wavenumbers are even less produced. Using the same technique for modes that become
nonrelativistic during the radiation era, $k_e < k < a_{\rm eq} m$,
we obtain
\be
\tilde\beta_k \simeq \frac{1}{4} \int_{-\infty}^{\infty} \frac{du}{1+u^2} 
e^{-i \frac{m_\psi k^2}{H_I k_e^2} [ u \sqrt{1+u^2} + {\rm ArcSinh}(u) ]} ,
\label{eq:tilde-beta-u-fermions-high-k}
\ee
which gives $\tilde\beta_k \simeq (\pi/6) e^{-\pi m_\psi k^2 / ( 2 H_I k_e^2 )}$,
from the contribution of the saddle-point at $u \simeq -i$.
Then, cutting the integral (\ref{eq:rho-tilde-beta}) at $k < k_e$ we obtain
\be
\rho(t_0) \simeq \frac{2 m_\psi^4}{27} e^{-2\pi m_\psi/H_I} 
\left( \frac{a(t_e)}{a_0} \right)^3 .
\ee
This gives
\be
\frac{\rho(t_0)}{\rho_{c0}} \simeq 10^{-3} \left( \frac{m_\psi}{H_I} \right)^4
e^{-2\pi m_\psi/H_I} \left( \frac{H_I}{10^7 \, {\rm GeV}} \right)^{5/2} ,
\label{eq:rho-m-large-HI}
\ee
which quickly becomes very small when $m_\psi \gg H_I$.

\subsubsection{Constraints on the model}

We have seen in Eq.(\ref{eq:L-psi-j-n-M}) that the compactification over $S^1$ leads to
the Kaluza-Klein towers of fermion with mass $M_{j,n}$ given in Eq.(\ref{eq:Mjn}).
For the zero-mode fermions with $n=0$ and $\alpha_j=0$ we require that $m_j$
obeys the upper bound (\ref{eq:mpsi-upper}) or is above $H_I$ as in Eq.(\ref{eq:rho-m-large-HI}).
For the higher modes $|n| \geq 1$ and for the zero-mode of the fermions with $\alpha_j \neq 0$
we require $1/R \gg H_I$.
This gives a negligible production for the full Kaluza-Klein tower thanks to the exponential
cutoff (\ref{eq:rho-m-large-HI}), with $M_{j,n} \sim |n|/R$ for large $|n|$.
Therefore, the gravitational production of the fermions is negligible and consistent with
observational data for models with
\be
m_j \ll 10^8 \, {\rm GeV} \;\; \mbox{if} \;\; m_j < H_I , \;\;\; \mbox{or} \;\; m_j \gg H_I , 
\label{eq:mj-2-options}
\ee
and
\be
1/R \gg H_I .
\label{eq:R-H_I-sup}
\ee
This last condition was already encountered in (\ref{eq:R-HI}).
The two cases (\ref{eq:mj-2-options}) are allowed by the conditions on the model parameters
derived in the previous sections.
Therefore, a wide range of fermion masses is permitted.

\subsection{Dark photons production}
\label{sec:Photon-prod}

In addition to the fermions considered in section~\ref{sec:Fermion-prod},
the integration over the fifth dimension gives rise to a tower of Kaluza-Klein photons 
$A_\mu^{(0)}$, $A_\mu^{(c,n)}$ and $A_\mu^{(s,n)}$ in the 4D Lagrangian
(\ref{eq:L-4D-tot}).
The background $A_\mu=0$ is a classical solution, 
together with $\psi_j^{(n)} = 0$.
However, the cosmological expansion of the Universe again gives rise to a time dependence
of the quantum vacuum, which leads to a gravitational production of particles, as for the
Kaluza-Klein fermions studied in section~\ref{sec:Fermion-prod}.
The zero mode $A_\mu^{(0)}$ is massless. Therefore it is conformally coupled to gravity, which
prohibits any gravitational production \cite{Kolb:2020fwh,Kolb:2023ydq}.
As seen in Eq.(\ref{eq:L-Amu}), the higher modes have a large mass $m_n = n/R$.
As for the fermions, with the exponential cutoff (\ref{eq:fermion-exp-cutoff}) and the condition
(\ref{eq:R-H_I-sup}, their production will be negligible if this mass is greater than the Hubble
expansion rate during inflation.
We recall here that we recover the same exponential cutoff (\ref{eq:fermion-exp-cutoff}) 
for these Kaluza-Klein massive dark photons.

We describe in appendix~\ref{app:production-photons} the formalism that we use to compute
their relic energy density (\ref{eq:rho-photons}), following \cite{Kolb:2020fwh,Kolb:2023ydq}.
This is obtained from the Bogolyubov coefficient $\beta_k$, defined in Eq.(\ref{eq:photons-in-out}),
associated with the change of vacuum between early and late times in the cosmological evolution.
From Eq.(\ref{eq:tilde-beta-photons}), it is sufficient to compute the late-time value of the
coefficient $\tilde\beta_k$, associated with the adiabatic representation (\ref{eq:photons-adiabatic})
\cite{Kofman:1997yn},
which obeys the equation of motion (\ref{eq:motion-adiabatic-photons}).
As in section~\ref{sec:fermions-high-mass}, the analysis simplifies in the high-mass limit
where the particle production is negligible.
Then, the coefficient $\tilde\alpha_k$ remains close to unity at all times while $\tilde\beta_k$
remains very small and the second equation (\ref{eq:motion-adiabatic-photons})
can be integrated as
\be
\tilde\beta_k \simeq \int d\eta \frac{\omega_k'}{2\omega_k} e^{-2i\Phi_k} .
\ee
During the inflationary stage (\ref{eq:Inflation-a}), we obtain at leading order in
$m_n \gg H_I$, for both transverse and longitudinal modes,
\be
\Phi_k \simeq - \frac{m_n}{H_I} \left( \sqrt{1+u^2} - {\rm ArcCoth} \sqrt{1+u^2} \right) ,
\ee
where we again made the change of variable $u=H_I k \eta/m_n$.
As for Eq.(\ref{eq:tilde-beta-u-fermions}), a saddle-point analysis gives the same exponential
cutoff as in (\ref{eq:fermion-exp-cutoff}),
\be
k < k_e : \;\; \tilde\beta_k \sim e^{-\pi m_n/H_I} ,
\label{eq:photon-exp-cutoff}
\ee
see also \cite{Kolb:2020fwh}.
Here we again introduced $k_e=a(t_e) m_n$ and modes at higher wavenumbers are further suppressed.
Since the Kaluza-Klein photons with $n \geq 1$ have a mass $m_n=n/R$, 
we conclude that they are not produced in any significant manner as long as $1/R \gg H_I$.
Thus, we recover the condition (\ref{eq:R-H_I-sup}).

\section{Theoretical self-consistency}
\label{sec:theoretical-consistency}

In our derivation of the dark matter potential in Section~\ref{sec:dark-matter-potential},
we neglected the contribution $V_I$ of Eq.(\ref{eq:V-I-1}) and we neglected the spacetime
dependence of $\phi$ in the mass term $M_{j,n}$ in the determinant (\ref{eq:V-eff-def}).
We now check that these approximations are valid.

\subsection{Validity of the derivative expansion}
\label{sec:derivative}

To derive the dark matter potential (\ref{eq:Va-Li}), we computed in Appendix~\ref{app:KK-modes}
the determinant or logarithmic trace (\ref{eq:L-4D-Tr-ln}) associated with the fermion
Lagrangian (\ref{eq:L-psi-j-n}) by neglecting the spacetime dependence of the dark matter
scalar field $\phi(x)$.
This is the lowest order in a derivative expansion, where we simply substitute in the
constant-field result (\ref{eq:V-a}) the local value of $\phi(x)$.
We compute in Appendix~\ref{app:derivative-expansion} the first corrections
due to the spacetime dependence of $\phi(x)$, taking into account the impact of nonzero
gradients $\partial_\mu\phi$ and neglecting higher-order derivatives.
From Eq.(\ref{eq:V-kappa-star}), the potential (\ref{eq:V-a}) is modified as
\ba
V & = & \sum_{j,n} \frac{1}{8\pi^2} \int_0^\infty \frac{ds}{s^3} \, 
e^{-s m_j^2 -s \zeta_{j,n}^2 \tanh(s\partial\zeta_{j})/(s\partial\zeta_{j})}  
\nonumber \\
&& \times \left( 1 - \frac{s^2 \partial\zeta_{j}^2}{2} \right) 
\sqrt{\frac{2 s \partial\zeta_{j}}{\sinh(2 s \partial\zeta_{j})}} ,
\label{eq:V-dzeta}
\ea
where we introduced the local amplitude of the gradient,
\be
\partial\zeta_{j} = \sqrt{ \partial_\mu \zeta_{j,n} \partial^\mu \zeta_{j,n} } 
\sim \frac{\partial\phi}{fR} ,
\ee
which does not depend on $n$.
We recover the zeroth-order approximation (\ref{eq:V-a}) for $\partial\zeta_{j} \to 0$.
For $s\to 0$ the correction terms $s^2\partial\zeta_{j}^2$ give rise to a logarithmic
UV divergence, as in the third term in (\ref{eq:ln-UV}), which can be removed by adding
a counterterm to the potential.
We can see in Eq.(\ref{eq:V-dzeta}) that
the local approximation (\ref{eq:V-a}) is valid as long as
$s \partial\zeta_{j} \ll 1$.
From Eq.(\ref{eq:V-s-7/2}), the integration in $s$ is exponentially damped for 
$s \gtrsim \max(1/m_j^2,\ell \pi R/m_j)$, where we assumed $m_j \neq 0$.
Therefore, the local approximation (\ref{eq:V-a}) is valid for
\be
\partial \phi \ll m_j f \, \min( 1 , m_j R ) .
\ee
In the nonrelativistic regime after the field has started oscillating we have
$\partial\phi \sim \sqrt{\rho_\phi}$, as seen from Eq.(\ref{eq:phi-cos-m}) below
and $\partial \phi \sim \partial_0 \phi \sim m \phi$.
Therefore, we obtain the constraint
\be
\sqrt{\rho_\phi} \ll m_j f \, \min( 1 , m_j R ) .
\label{eq:rho-phi-mjR}
\ee
Taking $m_1 \sim m_2 \sim m_\psi$ as in Eq.(\ref{eq:m1-m2-mpsi}) and using 
Eq.(\ref{eq:Omega-phi-f-m}) to express $f$ in terms of $m$, the condition
(\ref{eq:rho-phi-mjR}) evaluated at the time the scalar field starts to oscillates,
$H(t_{\rm osc}) \simeq m/3$, reads
\be
m_\psi \, \min(1, m_\psi R) \gg 10^{13} m .
\label{eq:derivative-condition}
\ee
For any ultralight dark matter mass $m < 1$ eV, there is a wide range of fermion
masses that satisfy the condition (\ref{eq:derivative-condition}).
The product $m_\psi R$ can also span a broad range and be much smaller or greater than
unity.
Therefore, the dark matter potential (\ref{eq:Va-Li}) is valid for a large range of
parameters $(m_\psi,R)$ and dark matter masses $m < 1$ eV.
The lower bound (\ref{eq:derivative-condition}) implies that the dark fermion masses
$m_j$ cannot be zero. Thus, the potential (\ref{eq:Va-m1=0}) and the curves
labeled $m_1=m_2=0$ in Figs.~\ref{fig:q2} and \ref{fig:Va} must be understood
as illustrative of the regime $m_\psi R \ll 1$.

\subsection{Bounds on the UV cutoff $\Lambda$}
\label{sec:cutoff-Lambda}

In Section~\ref{sec:dark-matter-potential}, we neglected the contribution $V_I$ of 
Eq.(\ref{eq:V-I-1}) to the dark matter potential.
Using the scalings (\ref{eq:V-I-Lambda}) and Eqs.(\ref{eq:Mjn}) and (\ref{eq:zeta-def})
we obtain
\be
\frac{\partial^2 V_I}{\partial \phi^2} \sim \frac{\Lambda^2+\zeta^2}{R^4 f^4} 
\sim \frac{\Lambda^2+1/R^2}{R^4 f^4} ,
\ee
where we wrote $\zeta_{j,n} \lesssim 1/R$ for $\phi \lesssim f$ from Eq.(\ref{eq:zeta-def}).
We assume the cutoff $\Lambda$ to be higher than $1/R$, so that the 5D theory
(\ref{eq:L-5D}) is valid. Then, $\frac{\partial^2 V_I}{\partial \phi^2} \sim \frac{\Lambda^2}{R^4 f^4}$.
Comparing with the contribution from $V_0$, which scales from Eq.(\ref{eq:m2-lambda4})
as $\frac{\partial^2 V_0}{\partial\phi^2} \sim \frac{1}{R^4 f^2}$ for $R m_{\psi}$ not too large,
we obtain the conditions
\be
\frac{1}{R} \ll \Lambda \ll f .
\label{eq:Lambda-bounds}
\ee
This range is nonzero because we have already required the condition $fR\gg 1$ in
(\ref{eq:fR-Rm_psi}).
This also ensures $\Lambda \gg H_I$ from Eq.(\ref{eq:R-HI}), so that our model applies during
the inflationary era.
Therefore, neglecting $V_I$ in Section~\ref{sec:dark-matter-potential} was justified,
taking a cutoff $\Lambda$ in the range (\ref{eq:Lambda-bounds}).

\section{Parameter space}
\label{sec:parameter-space}

We can now summarize the constraints obtained in the previous sections.
The parameters of our model (\ref{eq:L-5D}) are the 5D fermions masses $m_j$ and charges
$q_j$, the coupling $g_5$ and the radius $R$ of the fifth dimension,
\be
m_1, \;\; m_2 , \;\; q_1, \;\; q_2 , \;\; g_5, \;\; R .
\ee
The charges $q_j$ can be taken of the order of unity and the parameter space associated with repulsive
self-interactions was shown in Fig.~\ref{fig:q2}.
The fifth dimension radius $R$ is constrained by the small-scale gravity tests (\ref{eq:gravity-test}) and
the compactification during the inflationary stage (\ref{eq:R-EW}).
The latter condition also ensures that the Kaluza-Klein dark fermions and dark photons are not
appreciably produced during and after the inflationary stage, as seen in Eqs.(\ref{eq:R-H_I-sup})
and (\ref{eq:photon-exp-cutoff}).
The massless dark photon is not produced because it is conformally coupled, whereas the
zero-mode dark fermions are not produced provided their masses satisfy the condition
(\ref{eq:mj-2-options}), which permits a wide range of masses.
Within the misalignment mechanism, the dark matter mass $m$ is related to the fifth dimension 
radius $R$ by Eq.(\ref{eq:R-m}). 
We can see that all values in the range $10^{-21} {\rm eV} < m < 1 \, {\rm eV}$
lead to allowed values of $R$, for both low or high fermion masses $m_\psi$.
The isocurvature bounds give the upper bound (\ref{eq:HI-iso}) on the inflation scale,
which again permits all values of the dark matter mass in the range $10^{-21} {\rm eV} < m < 1 \, {\rm eV}$.

From the point of view of the dark matter dynamics, the relevant parameters are the dark matter
mass $m$ and the coupling $\lambda_4$ in Eq.(\ref{eq:m2-lambda4}).
We have seen above that $m$ can take any value in the range $10^{-21} {\rm eV} < m < 1 \, {\rm eV}$.
The coupling $\lambda_4$ is then determined by the misalignment mechanism (\ref{eq:lambda4-m}),
which can be expressed in terms of the characteristic density $\rho_a$ in Eq.(\ref{eq:rho_a-m})
or of the soliton radius $R_s$ in Eq.(\ref{eq:Rs-pc-2}).

We may also comment on possible coincidence and fine-tuning issues.
First, this model does not address the coincidences between the dark matter density and the baryon
or dark energy densities today, $\Omega_{\rm DM} \sim \Omega_b \sim \Omega_\Lambda$.
The observed dark matter density is simply reproduced by choosing the decay constant $f$ as 
a function of $m$ in Eq.(\ref{eq:f-abundance}), or equivalently the fifth dimension radius $R$
as in Eq.(\ref{eq:R-m}), for a generic initial angle $\theta_i \sim 1$. Thus, no additional
coincidence is introduced but none is explained either.
Second, the hierarchies between scales, $m \ll 1/R \ll f$, do not require a specific tuning
between the fermion parameters, because the Scherk-Schwarz twists make repulsive self-interactions 
generic and the dark matter mass $m$ is strongly suppressed in Eq.(\ref{eq:m2-lambda4})
by the exponential factor $e^{-2\pi R m_\psi}$ for a moderate value of $R m_\psi$
and by the large product $f R \gg 1$.
The hierarchy $f R \sim 10^{13}$ in Eq.(\ref{eq:fR}) (for $m \sim 10^{-6} \, {\rm eV}$) 
is equivalent to a small 4D gauge coupling, 
$g_4 = g_5/\sqrt{2\pi R} = 1/(2\pi f R) \sim 10^{-14} (m/10^{-6} \, {\rm eV})^{5/8}$.
This is stable under radiative corrections, since $g_4 \to 0$ restores a global symmetry, 
but its value is not explained by the model.

\begin{figure}
\centering
\includegraphics[height=6.5cm,width=0.45\textwidth]{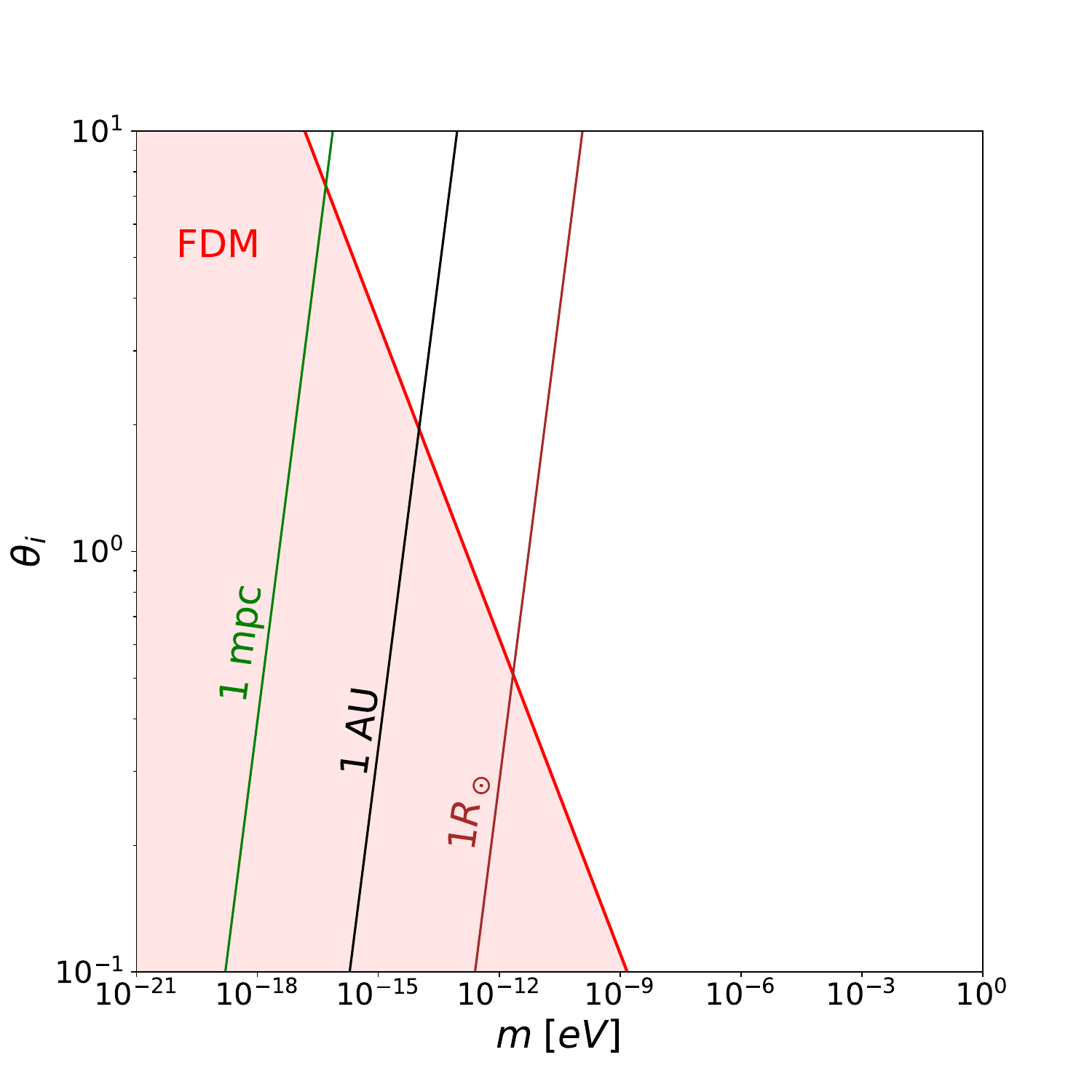}
\caption{
The red shaded region labeled ``FDM'' is the domain in the $(m,\theta_i)$ plane 
where the quantum pressure dominates over the repulsive self-interaction.
The almost vertical lines correspond to a soliton radius $R_s=1$ mpc, 1 AU and 1 $R_\odot$.
}
\label{fig:m-theta}
\end{figure}

We show in Fig.~\ref{fig:m-theta} the behavior of the dark matter solitons (\ref{eq:soliton-rho}),
in the plane $(m,\theta_i)$.
In the red shaded region, obtained from Eq.(\ref{eq:Rs-dB}) with $v=250$ km/s,
the quantum pressure dominates over the repulsive self-interactions and determines the
soliton profile.
In this regime, we recover a fuzzy dark matter model.
The almost vertical lines correspond to fixed values of the soliton radius $R_s$ in 
Eq.(\ref{eq:Rs-pc-2}), with $R_s=1$ mpc, 1 AU and 1 $R_\odot$, from left to right.
In these models, where the misalignment mechanism
determines the value of the self-interaction coupling $\lambda_4$ from Eq.(\ref{eq:lambda4-m}),
up to the initial angle $\theta_i$, the solitons are typically of astrophysical size and smaller than
globular clusters.
To reach much larger subgalactic size would require large values of $\theta_i$.
This may be possible if the field $\phi$ starts with a non-negligible kinetic energy, so that it can 
pass over many cycles of the potential (\ref{eq:Va-hn}) before settling down around the
final minimum at $\phi=0$.
However, this introduces some degree of fine tuning and we do not investigate further this scenario.

\section{Model with low-mass dark fermions $m_j R \ll 1$}
\label{sec:low-mass}

The case where both fermions $\psi_j$ of the 5D Lagrangian (\ref{eq:L-5D}) have a low mass,
$m_j R \ll 1$, requires additional care because, as seen in Fig.~\ref{fig:Va}, 
the minimum of the potential at $\phi=0$ with a repulsive quartic self-interaction
is only a local minimum. Therefore, we need to study whether the field can remain trapped
in the false vacuum during the cosmological evolution.
A second feature that is specific to this case is that the potential has formally an
infinite $\lambda_4$ quartic coupling constant in (\ref{eq:m1-m2-0}), because the potential
has a logarithmic singularity at $\phi=0$.
We investigate these two points in this Section.
As explained in Section~\ref{sec:derivative},
the condition (\ref{eq:derivative-condition}) implies that the masses $m_j$ are not zero
but they can be much smaller than $1/R$ and in the following we consider the massless limit
$m_1=m_2=0$, understood as the limit of the regime $m_j R \ll 1$.

\subsection{Metastability}

We first consider the metastability of the false vacuum at $\phi=0$ and estimate the
escape rate by instanton and sphaleron processes.
Instantons correspond to classical trajectories in Euclidean space connecting the
two vacua, i.e. interpolating between the false and true vacua in a metastable situation. A sphaleron is a classical trajectory which just overcomes the barrier between the false and true vacua given enough time and energy \cite{Mukhanov:2005sc}.
The potential barrier is of the same order as the difference between the local and global minima,
as the only two dimensional parameters are the amplitude of the potential, $1/(2 \pi R)^4$, and
the scale $f$ of the field $\phi$ (for a charge $q_2$ of the order of unity).
This is beyond the strict validity of the thin-wall approximation.
However, this still provides the correct scaling in terms of $R$ and $f$.

In the thin-wall approximation, for a potential with a false vacuum at $\phi=0$, with
$V(0)=0$, and a true vacuum at $\phi_0$, with $V(\phi_0)= - \epsilon$, 
one defines the quantity $\sigma$ by \cite{Mukhanov:2005sc}
\be
\sigma = \int_0^{\phi_0} d\phi \, \sqrt{2V} .
\label{eq:sigma-def}
\ee
The true vacuum bubble nucleation rate $\Gamma$ is the maximum 
of the rates associated with instantons,
\be
\Gamma_I \sim e^{-S_I} , \;\;\;  S_I = \frac{27 \pi^2 \sigma^4}{2\epsilon^3} ,
\ee
and with sphalerons,
\be
\Gamma_{\rm sph} \sim e^{-V_{\rm sph}/T} , \;\;\; V_{\rm sph} = \frac{16 \pi \sigma^3}{3 \epsilon^2} ,
\ee
where we discarded the prefactors.
In our case, we take $\phi_0 = 2 \pi f$ and $\sigma = \phi_0 \sqrt{2 V_0}$ with $V_0 = 1/(2\pi R)^4$.
This gives
\be
\Gamma_I \sim e^{-(24 \pi^4)^3 (Rf)^4} , \;\;\;  \Gamma_{\rm sph} \sim e^{- 2^{10} \sqrt{2} \pi^7 R^2 f^3/(3T)} .
\ee
Using Eq.(\ref{eq:R-m}), with $m_\psi=0$, and Eq.(\ref{eq:f-abundance}), we obtain
\be
\ln\Gamma_I \sim - 10^{62} \theta_i^{-1} \left( \frac{m}{10^{-6} \, {\rm eV}} \right)^{-5/2} ,
\ee
and
\be
\ln\Gamma_{\rm sph} \sim - 10^{36} \theta_i^{-3/2} 
\left( \frac{m}{10^{-6} \, {\rm eV}} \right)^{-3/2} 
\left( \frac{H_I}{1 \, {\rm GeV}} \right)^{-1/2} .
\ee
Therefore, the escape rate from the false vacuum is very small, both for the instanton
and sphaleron processes, and the scalar field remains trapped in the local minimum around
$\phi \simeq 0$ until today.

\subsection{Low-energy potential}

We derived in Eq.(\ref{eq:Va-m1=0}) the potential of the dark matter pseudoscalar $\phi$
in the case $m_1=m_2=0$.
Expanding up to quartic order in $\phi$, it reads as
\be
|\phi| \ll f : \;\;\; V(\phi) = \frac{m^2}{2} \phi^2 + \frac{\lambda_4}{4} \phi^4 
\left[ \alpha - \beta \ln (\phi^2/f^2) \right] ,
\label{eq:V-ln-alpha-beta}
\ee
where $m_2$ was given in Eq.(\ref{eq:m1-m2-0}) and we introduced
\be
\lambda_4 = \frac{1}{384 \pi^6 R^4 f^4} , \;
\alpha = 25 q_2^4 -12 \ln 2 -6 q_2^4 \ln q_2^2 , \; \beta = 6 q_2^4 .
\ee
In the nonrelativistic regime, it is convenient to write the scalar field as
\be
\phi = \frac{\sqrt{2 \rho}}{m} \cos(m t - S) ,
\label{eq:phi-cos-m}
\ee
where $\rho(\vec x,t)$ is the dark matter density and $S(\vec x,t)$ is the phase,
which defines the curl-free velocity field $\vec v = \vec\nabla S/m$.
As $\rho$ and $S$ evolve on cosmological or astrophysical timescales, we can average
over the fast oscillations of angular frequency $m$. Substituting into the quartic part
of the potential (\ref{eq:V-ln-alpha-beta}) this averaging gives at leading order
the interacting potential
\be
\rho \ll \rho_a : \;\;\; V_I(\rho) = \frac{\rho^2}{2\rho_a} \ln(\rho_a/\rho) ,
\ee
with
\be
\rho_a = \frac{4 \zeta(3) (3-4 q_2^2)}{q_2^4} m^2 f^2 .
\label{eq:rho_a-m2f2}
\ee
This also determines the potential $\Phi_I$ as
\be
\Phi_I(\rho) = \frac{dV_I}{d\rho} \simeq \frac{\rho}{\rho_a} \ln(\rho_a/\rho) .
\ee
This is the potential that enters the Schr\"odinger equation that governs the dynamics
in the nonrelativistic regime or the Euler equation in the hydrodynamical mapping.
In particular, in the regime $R_s \gg \lambda_{\rm dB}$ where the self-interaction pressure
dominates over the quantum pressure, the soliton profile is the solution of the equation of
hydrostatic equilibrium $\vec\nabla (\Phi_N+\Phi_I) = 0$, where $\Phi_N$ is the Newtonian
gravitational potential.
As compared with Eqs.(\ref{eq:PhI-rho_a}) and (\ref{eq:Rsol-def}), we can see that at leading
order the soliton radius exhibits a weak logarithmic dependence on the typical density
of the system
\be
R_s = \sqrt{\frac{\pi}{4 {\cal G} \rho_a}} \sqrt{ \ln(\rho_a/\rho) }  .
\label{eq:Rsol-rho}
\ee
For a density of the order of the dark matter density in the solar system,
$\rho \sim 0.4 \, {\rm GeV/cm^3}$, we obtain $\sqrt{ \ln(\rho_a/\rho) } \sim 9.6$ and
\be
R_s \sim 0.17 \, {\rm pc} \, \theta_i \left( \frac{m}{10^{-6} \, {\rm eV}} \right)^{-3/4} .
\ee
Thus, in spite of the weak logarithmic dependence we recover the same length scale as for
the generic case of massive fermions in Eq.(\ref{eq:Rs-pc-2}).

As a result, the case of low fermion masses can be considered on par with other ranges of masses. 
The large amplitude of the potential barrier guarantees the stability of the origin in field space. Moreover, the case of strictly massless fermions hardly modifies the shape of the soliton profiles.

\section{Conclusion}
\label{sec:Conclusion}

In this paper, we have constructed 5D models of scalar dark matter with repulsive self-interactions. 
This leads to the formation of solitons with flat cores where gravity is balanced by the effective
pressure due to the self-interactions, rather than the quantum pressure due to wavelike effects
as in fuzzy dark matter.
These models rely on the coupling of a five-dimensional $U(1)$ gauge field to at least two massive 
fermions in 5D. 
The dark matter is represented by the fifth component of the gauge field and its potential is obtained by 
integrating out the massive fermions. The parameter space of the model is enlarged by the Scherk-Schwarz 
twists of the fermions around the fifth dimension. This makes the range of parameters leading to repulsive 
interactions relatively generic and avoids fine-tunings amongst the parameters of the model. 

The parameters of the model are the radius $R$ of the compact fifth dimension, the masses $m_j$
and charges $q_j$ of the fermions and their coupling constant $g_5$ to the $U(1)$ gauge field.
The charges can be taken of the order of unity while the masses $m_j$ can span a wide range,
from small to large values as compared with $1/R$.
This mostly leaves two parameters, the fifth-dimension radius $R$ and the coupling $g_5$,
which can be expressed in terms of a decay constant $f$ that determines the period of the dark matter
potential.
These two parameters, $R$ and $f$, can be traded for the mass $m$ of the dark matter pseudoscalar
and its quartic coupling constant $\lambda_4$.
A third final parameter is the initial misalignment $\theta_i$ of the scalar field, at the start of the
oscillating phase when the Hubble expansion rate falls below $m$. 

The dark matter abundance depends on powers of $\theta_i$, $f$ and $m$.
Then, for typical angles $\theta_i \sim 1$, the measured value of $\Omega_m$ determines
$f$, or $\lambda_4$, in terms of $m$, which is the single remaining free parameter.
As explained above, the fermions masses $m_j$ are additional parameters but they do not directly
impact the low-energy phenomenology.
For large enough masses $m\gtrsim 10^{-12}$ eV, generic misalignment angles $\theta_i \lesssim 1$
lead to soliton radii of astrophysical size and smaller than globular clusters.
They would be too small to address the core-cusp issue of CDM but could
compose a new form of dark matter distributed on small scales and change the properties of dark matter
at short distance. This could either compose parts or the entire dark matter of the Universe in a similar 
fashion as primordial black holes could be parts or the whole dark matter that we observe.
Larger soliton sizes, up to $1 \,{\rm pc}$, require large misalignment angles (which may be associated
with large initial kinetic energy). Larger sizes, up to $1 \,{\rm kpc}$ are possible at low scalar mass,
but in such cases the self-interactions are negligible and we recover the usual fuzzy dark matter
behaviors.
 
In all these cases, we have considered that compactification takes place before inflation implying that the 
potential formation of scalar domain walls is suppressed. 
This also allows us to check that the model is consistent with theoretical and observational constraints
from the inflation era to the current time. 

The small number of ingredients and parameters in this model, with a common origin to the mass and
self-interactions of the pseudoscalar $\phi$, i.e. the one-loop potential generated by the charged
5D fermions, leads to a unique relation between the quartic coupling $\lambda_4$ and the mass $m$,
up to a factor $\theta_i^2$.
Generic misalignment $\theta_i \lesssim 1$ then gives an upper bound on the soliton size.
This could be evaded by different initial conditions (corresponding to large $\theta_i$),
for instance due to kicks induced by a quadratic coupling to the standard model \cite{Burrage:2026pei}.
Alternatively, more complex models where the dark matter potential combines different contributions,
as in monodromy models, and decouples $\lambda_4$ from $m$, could give rise to a wider range
of soliton sizes and different phenomenology. This is left for future work.

The variety of phenomena associated with self-interacting scalar dark matter, such as the formation
and merging of solitons, wavelike effects and vortices \cite{Brax:2025uaw,Brax:2025vdh},
deserve further investigations. In particular, numerical simulations where the collective behavior
of the solitons of various sizes and their role in the dynamics of dark matter on large and small scales
would be welcome. We hope to come back on these issues in the future. 

In this paper, we have focused on the dark sector, taking into account self-interactions but
assuming vanishing couplings with the standard model (except of course with gravity).
This model could be extended by studying possible embeddings of the dark matter within the standard 
model of particle physics. In particular, the interactions between the dark matter field and ordinary matter 
should be restricted by gauge invariance. This could be relevant in view of the swath of experiments 
\cite{Arza:2026rsl} looking for dark matter in the laboratory via its manifestation in atomic phenomena 
such as clocks or interferometry \cite{Hees:2018fpg}. The study of these couplings is left for future work.

\appendix

\section{Integration over the fermion KK modes}
\label{app:KK-modes}

In this appendix, we compute the potential (\ref{eq:L-4D-Tr-ln}), which arises from the
integration over the fermions.
We can consider the various fermions separately and we first define for each contribution
\be
V = i \, {\rm Tr} \left[ \langle x | \ln \left( i \gamma^\mu \partial_\mu - m  \right) | x \rangle \right] ,
\label{eq:L-a-m}
\ee
where $m$ is constant. Differentiating with respect to $m$ gives
\cite{Schwartz_2013},
\ba
\frac{\partial V}{\partial m} & = & - i \, {\rm Tr}  \langle x | \frac{1}{i \gamma^\mu \partial_\mu - m} | x \rangle
= i \, {\rm Tr}  \langle x | \frac{i \gamma^\mu \partial_\mu + m}{\partial_\mu \partial^\mu + m^2} | x \rangle
\nonumber \\
&& = 4 i m \, \langle x | \frac{1}{\partial_\mu \partial^\mu + m^2} | x \rangle .
\label{eq:dV-dm-1}
\ea
The fraction can be written in the integral form as \cite{Schwartz_2013},
\be
\frac{\partial V}{\partial m} = - 4 m \int_0^{\infty} ds \, e^{-i s m^2 - \epsilon s}
\langle x | e^{-i s \partial_\mu \partial^\mu} | x \rangle ,
\label{eq:dVdm-1}
\ee
with $\epsilon \to 0^+$.
Integrating over $m$ gives
\ba
V & = & - 2 i  \int_0^{\infty} \frac{ds}{s} \, e^{-i s m^2 - \epsilon s}
\int \frac{d^4k}{(2\pi)^4} e^{i s k^2} \nonumber \\
& = & - \frac{1}{8\pi^2}  \int_0^{\infty} \frac{ds}{s^3} \, e^{-i s m^2 - \epsilon s} ,
\label{eq:V-d4k}
\ea
where we discarded an irrelevant integration constant.
It is implicitly understood that the divergence at $s \to 0$ is regularized by counterterms, as in
a Pauli-Villars renormalization scheme.
Rotating $s \to - i s$ we obtain
\be
V = \frac{1}{8\pi^2}  \int_0^{\infty} \frac{ds}{s^3} \, e^{- s m^2} ,
\label{eq:L-a-m-2}
\ee
which leads to Eq.(\ref{eq:V-a}).

\section{Derivative expansion}
\label{app:derivative-expansion}

In this appendix we consider the corrections to the dark matter potential (\ref{eq:Va-Li})
due to the spacetime dependence of $\phi$ that was neglected in the computation of the determinant
from (\ref{eq:L-4D-Tr-ln}).
If we go beyond the approximation of constant $\phi$, we cannot absorb the phase of the complex mass
$m+i\zeta \gamma^5$ in (\ref{eq:L-psi-j-n}) by a global phase shift.
Then, from the expression (\ref{eq:L-psi-j-n}) we obtain the potential associated with the
Kaluza-Klein fermion mode $\psi_j^{(n)}$,
\be
V(\phi) = i \, {\rm Tr} \left[ \langle x | \ln \left( i \gamma^\mu \partial_\mu - m 
- i \zeta \gamma^5 \right) | x \rangle \right] .
\label{eq:L-Tr-ln-zeta}
\ee
In a fashion similar to the constant-$\phi$ case (\ref{eq:dV-dm-1}),
taking a derivative with respect to $m$ gives
\be
\frac{\partial V}{\partial m} = i \, {\rm Tr} \langle x | \frac{i \gamma^\mu \partial_\mu + m 
- i \zeta \gamma^5}{\partial_\mu \partial^\mu + m^2 + \zeta^2 
- \gamma^\mu \gamma^5 \partial_\mu \zeta } | x \rangle ,
\ee
which we again write as \cite{Schwartz_2013}
\ba
\frac{\partial V}{\partial m} & = & - \int_0^\infty ds \, e^{-i s m^2 -\epsilon s} \,
{\rm Tr} \langle x | e^{-i s ( \partial_\mu \partial^\mu + \zeta^2 
- \gamma^\mu \gamma^5 \partial_\mu \zeta ) } \nonumber \\
&& \times ( i \gamma^\mu \partial_\mu + m - i \zeta \gamma^5 ) | x \rangle .
\ea
In the following, we focus on the first correction in a derivative expansion and we
take $\partial_\mu \zeta$ constant and small. Then, we can separately expand the exponential term
$e^{i s \gamma^\mu \gamma^5 \partial_\mu \zeta}$ and take the trace over the $\gamma$ matrices
of each power multiplied by the factor $( i \gamma^\mu \partial_\mu + m - i \zeta \gamma^5 )$.
Using the properties of traces of products of $\gamma$ matrices \cite{Peskin:1995ev,Schwartz_2013}, 
${\rm Tr}[{\rm odd } \; \# \; {\rm of} \; \gamma \; {\rm matrices}]=0$ and
${\rm Tr}[\gamma^\mu \gamma^\nu \gamma^5]=0$,
we find that the lowest-order
nonzero term is quadratic in $\partial_\mu\zeta$ and we obtain at lowest order
\ba
s \partial_\nu \zeta \ll 1 : \;\;\;
\frac{\partial V}{\partial m} & = & - 4 m \!\! \int_0^\infty \!\!\!  ds \, e^{-i s m^2 -\epsilon s} \,
\left( \! 1 \! + \! \frac{s^2}{2} \partial_\nu \zeta \partial^\nu \zeta \! \right) \nonumber \\
&& \times \langle x | e^{-i s ( \partial_\mu \partial^\mu + \zeta^2 )} | x \rangle ,
\ea
which agrees with Eq.(\ref{eq:dVdm-1}) for constant $\zeta$ (and $m^2 \to m^2+\zeta^2$).
Integrating over $m$ gives
\ba
V & = & - 2 i \int_0^\infty \frac{ds}{s} \, e^{-i s m^2 -\epsilon s} \,
\left( 1 + \frac{s^2}{2} \partial_\nu \zeta \partial^\nu \zeta \right) \nonumber \\
&& \times \langle x | e^{-i s ( \partial_\mu \partial^\mu + \zeta^2 )} | x \rangle .
\label{eq:V-s-dzeta}
\ea
As compared with Eq.(\ref{eq:V-d4k}), the eigenfunctions of the operator in the exponential
are no longer planes waves because $\zeta$ is not constant.
Nevertheless, at lowest order in the derivative expansion $\zeta$ displays a linear dependence
on $x^\mu$ so that we have a simple harmonic oscillator, which can be exactly integrated
\cite{Fradkin_2021}.
Thus, expanding $\zeta(x)$ around a reference point $x_\star$, we write at linear order
\be
\zeta(x) = \zeta_\star + (x^\mu-x^\mu_\star) \partial_\mu\zeta_\star ,
\ee
and we need to compute the matrix element
\be
I = \langle y | e^{-i s [ \partial_\mu \partial^\mu + (\zeta_\star + (x^\mu-x^\mu_\star) \partial_\mu\zeta_\star )^2 ] } | x \rangle_{x=y=x_\star} .
\label{eq:I-linear-matrix}
\ee
Making the change of coordinate $x'=x-x_\star$, performing a Wick rotation, 
and next a rotation in Euclidian space with one direction $x_\parallel$
along the vector $\partial_\mu\zeta_\star$ and three transverse directions $\vec x_\perp$,
the matrix element $I$ only differs from the free field case (\ref{eq:V-d4k}) along the 
longitudinal direction. Therefore, we factor out the free case as
\be
I = \frac{-i}{16 \pi^2 s^2} \, e^{-i s \zeta_\star^2} \, \frac{J(\kappa_\star)}{J(0)} ,
\label{eq:I-J-def}
\ee
where we introduced the one-dimensional matrix element
\be
J(\kappa_\star) = \langle y | e^{-is \hat H} | x \rangle_{x=y=0} ,
\ee
with the one-dimensional Hamiltonian
\be
\hat H = \hat p^2 + (\zeta_\star + \kappa_\star \hat x)^2 , \;\;\;
\kappa _\star= | \partial_\mu\zeta_\star | = \sqrt{ | \partial_\mu\zeta_\star 
\partial^\mu\zeta_\star | } .
\ee
Making the change of variable $x'=x+\zeta_\star/\kappa_\star$ we recover the harmonic
oscillator
\be
J = \langle y | e^{-is \hat H} | x \rangle_{x=y=\zeta_\star/\kappa_\star} , \;\;\;
\hat H = \hat p^2 + \kappa_\star^2 \hat x^2 .
\ee
As usual \cite{Fradkin_2021}, this matrix element can be written as the Feynman path integral
\be
J = \int {\cal D} q \, e^{i S(q,\dot q)} , \;\;\; S = \int_0^s dt \left[ \frac{1}{4} \dot q^2 
- \kappa_\star^2 q^2 \right] ,
\ee
where the trajectories $q(t)$ satisfy the boundary conditions $q(0)=q(s)=q_\star$
with $q_\star = \zeta_\star/\kappa_\star$.
As the action is quadratic, this can be simplified by writing $q(t) = q_c(t)+\xi(t)$,
where $q_c(t)$ is the solution of the classical equation of motion and $\xi(t)$
satisfies the boundary conditions $\xi(0)=\xi(s)=0$, with
\be
J =  e^{i S(q_c,\dot q_c)} Z_2 , \;\;\; Z_2 = \int {\cal D} \xi \, e^{i S(\xi,\dot \xi)}
= ( \det \hat A )^{-1/2} , 
\label{eq:J-Z2}
\ee
with the operator $\hat A$,
\be
\hat A = - \frac{1}{2} \frac{d^2}{dt^2} - 2 \kappa_\star^2 , \;\;\;
S(\xi,\dot \xi) = \frac{1}{2} \int_0^s dt \, \xi \cdot \hat A \cdot \xi .
\ee
The eigenfunctions $\psi_n$ and eigenvalues $\lambda_n$ of $\hat A$ are
\be
\psi_n = \sqrt{2}{s} \sin(n \pi t/s) , \;\; \lambda_n = \frac{n^2 \pi^2}{2 s^2} - 2 \kappa_\star^2 ,
\;\; n = 1, 2, \dots
\ee
This gives \cite{Gradshteyn:2015}
\be
\frac{Z_2(\kappa_\star)}{Z_2(0)} = \left[ \prod_{n=1}^\infty \left( 1 - 
\frac{4 \kappa_\star^2 s^2}{n^2\pi^2} \right) \right]^{-1/2} = 
\sqrt{\frac{2\kappa_\star s}{\sin(2\kappa_\star s)}} .
\ee
The value of the action along the classical solution $q_c(t)$ reads
\be
S(q_c,\dot q_c) = - \kappa_\star q_\star^2 \tan(\kappa_\star s) .
\ee
Substituting into Eqs.(\ref{eq:I-J-def}) and (\ref{eq:J-Z2}), the matrix element 
(\ref{eq:I-linear-matrix}) reads
\be
I = \frac{-i}{16 \pi^2 s^2} \, \sqrt{\frac{2\kappa_\star s}{\sin(2\kappa_\star s)}}
\, e^{-i \zeta_\star^2 \tan(\kappa_\star s)/\kappa_\star} .
\ee
This gives for the potential (\ref{eq:V-s-dzeta})
\ba
V & = & - \frac{1}{8\pi^2} \int_0^\infty \frac{ds}{s^3} \, e^{-\epsilon s -i s m^2 
- i \zeta_\star^2  \tan(\kappa_\star s)/\kappa_\star} 
(1 + s^2 \kappa_\star^2/2) \nonumber \\
&& \times \sqrt{\frac{2\kappa_\star s}{\sin(2\kappa_\star s)}} .
\label{eq:V-kappa-star}
\ea
Performing a Wick rotation $s \to - i s$ we obtain Eq.(\ref{eq:V-dzeta}).

\section{Gravitational production of Kaluza-Klein dark fermions}
\label{app:production-fermions}

In this appendix, we compute the gravitational production of the fermions $\psi_j^{(n)}$
in the 4D Lagrangian (\ref{eq:L-4D-tot}), due to the expansion of the Universe.
As for primordial curvature fluctuations, the main production will be associated with the
crossing of the horizon by the modes during the inflationary stage.
We neglect metric perturbations and focus on the impact of the expansion of the homogeneous
FLRW background.
The Lagrangian (\ref{eq:L-psi-j-n-M}) gives in curved spacetime the action
\be
\mathcal{S} = \int d^4x \, \sqrt{-g} \left[ i \bar{\Psi} \tilde\gamma^\mu \nabla_\mu \Psi 
- m \bar{\Psi} \Psi \right],
\label{eq:S-Psi}
\ee
where we simplified notations and $m$ denotes the fermion mass in this appendix. 
The covariant derivative $\nabla_\mu$ includes the spin connection
and $\tilde\gamma^\mu$ are the curved spacetime counterparts of the usual Dirac matrices
$\gamma^\mu$ \cite{Birrell_Davies_1982}.
We work with the conformal time $\eta$, so that the FLRW metric reads
\be
ds^2 = a^2(\eta) \left( d\eta^2 - d\vec x^{\,2} \right) , \;\; dt = a(\eta) d\eta ,
\label{eq:FLRW}
\ee
where $a(\eta)$ is the scale factor, and the Dirac equation reads
\be
(i \gamma^\mu \partial_\mu - a m) \psi = 0 , \;\;\; \mbox{with} \;\;\; \psi = a^{-3/2} \Psi .
\label{eq:Dirac-eta}
\ee
Thus, one recovers the usual Dirac equation in terms of the rescaled field $\psi$, but with
a time-dependent mass.
This time dependence gives rise to a time-dependent vacuum and a production of particles.
In contrast with the case of a scalar field, it is clear that this time dependence and therefore
the particle production vanish in the massless limit $m\to 0$.
This is due to the fact that the action (\ref{eq:S-Psi}) is conformally invariant for
$m=0$ and the FLRW metric (\ref{eq:FLRW}) is a conformal transformation of the Minkowski metric
\cite{Birrell_Davies_1982,Kolb:2023ydq}.
Then, we can quantize the field $\psi$ in the canonical fashion
\be
\hat\psi = \sum_\lambda \int \frac{d\vec k}{(2\pi)^{3/2}} \left[ \hat{a}_{\vec k,\lambda} 
u_{\vec k,\lambda}(\eta) e^{i\vec k \cdot \vec x} + \hat{b}_{\vec k,\lambda}^{\dagger} 
v_{\vec k,\lambda}(\eta) e^{-i\vec k \cdot \vec x} \right] 
\label{eq:psi-modes-operators}
\ee
where $\lambda=\pm 1$ is the helicity, $\hat a$ and $\hat b$ are the ladder operators,
and $u_{\vec k,\lambda}(\eta)$ and $v_{\vec k,\lambda}(\eta)$ the mode functions that are solution
of the time-dependent Dirac equation (\ref{eq:Dirac-eta}).
The helicity dependence can be factorized as \cite{Chung:2011ck,Boyle:2018rgh,Kolb:2023ydq}
\be
u_{\vec k,\lambda}(\eta) = \left( \begin{array}{c} u_{A,k}(\eta) h_{\hat{k},\lambda} \\
\lambda u_{B,k}(\eta) h_{\hat{k},\lambda} \end{array} \right) , 
\ee
and
\be
v_{\vec k,\lambda}(\eta) = \left( \begin{array}{c} -u_{B,k}^\star(\eta) h_{-\hat{k},\lambda} \\
\lambda u_{A,k}^\star(\eta) h_{-\hat{k},\lambda} \end{array} \right) e^{-i\lambda\phi} ,
\ee
where $\hat k = \vec k/k$ is the unit vector along the direction $\vec k$. 
The two-component spinors $h_{\hat{k},\lambda}$ are eigenspinors of the helicity
operator $\frac{1}{2} \hat{k} \cdot \sigma$ with eigenvalue $\lambda = \pm 1$.
In spherical coordinates, with $\vec k = k (\sin\theta\cos\phi,\sin\theta\sin\phi,\cos\theta)$,
they read as
\be
h_{\hat{k},+1} = \left( \begin{array}{c} e^{-i\phi} \cos\frac{\theta}{2} \\ \sin\frac{\theta}{2} 
\end{array} \right) , \;\;
h_{\hat{k},-1} = \left( \begin{array}{c} e^{-i\phi} \sin\frac{\theta}{2} \\ -\cos\frac{\theta}{2} 
\end{array} \right) .
\ee
The creation and annihilation operators obey the anticommutation relations
\be
\{ \hat{a}_{\vec k,\lambda} , \hat{a}_{\vec k',\lambda'}^\dagger \} = 
\{ \hat{b}_{\vec k,\lambda} , \hat{b}_{\vec k',\lambda'}^\dagger \} = \delta_D(\vec k - \vec k') 
\delta_{\lambda,\lambda'} ,
\ee
all others vanishing, while the mode functions satisfy the normalization
\be
| u_{A,k}(\eta) |^2 + | u_{B,k}(\eta) |^2 = 1 .
\ee
The Dirac equation (\ref{eq:Dirac-eta}) becomes
\be
i \partial_\eta \left( \begin{array}{c} u_{A,k} \\ u_{B,k} \end{array} \right) = 
\left( \begin{array}{lr} a m & k \\ k & - a m \end{array} \right)
\left( \begin{array}{c} u_{A,k} \\ u_{B,k} \end{array} \right) .
\label{eq:motion-uA-uB}
\ee
It is convenient to factorize the Minskowski solution, which would give the leading order
solution for adiabatic evolution of the vacuum \cite{Kofman:1997yn}, by writing the mode functions as
\cite{Kolb:2023ydq}
\be
\left( \! \begin{array}{c} u_{A,k} \\ u_{B,k} \end{array} \! \right) = 
\left( \begin{array}{lr} \sqrt{\frac{\omega_k+am}{2\omega_k}} e^{-i\Phi_k} & 
- \sqrt{\frac{\omega_k-am}{2\omega_k}} e^{i\Phi_k} \\ \sqrt{\frac{\omega_k-am}{2\omega_k}} 
e^{-i\Phi_k} & \sqrt{\frac{\omega_k+am}{2\omega_k}} e^{i\Phi_k} \end{array} \right) 
\left( \! \begin{array}{c} \tilde\alpha_k \\ \tilde\beta_k \end{array} \! \right) ,
\label{eq:tilde-alpha-tilde-beta}
\ee
with
\be
\omega_k(\eta) = \sqrt{k^2+a^2 m^2} , \;\; \Phi_k(\eta) = \int^\eta d\eta' \omega_k(\eta') .
\label{eq:omega-k}
\ee
The coefficients $\tilde\alpha_k(\eta)$ and $\tilde\beta_k$ satisfy the normalization
$| \tilde\alpha_k |^2 + | \tilde\beta_k |^2 = 1$ and Eq.(\ref{eq:motion-uA-uB})
gives for their equations of motion \cite{Kolb:2023ydq}
\be
\tilde\alpha_k' = - \frac{a' m k}{2 \omega_k^2} e^{2 i \Phi_k} \tilde\beta_k , \;\;\;
\tilde\beta_k' = \frac{a' m k}{2 \omega_k^2} e^{-2 i \Phi_k} \tilde\alpha_k .
\label{eq:tilde-alpha-beta}
\ee
For a slow cosmological evolution, $a' \to 0$, the two coefficients decouple and become constant.
This regime also applies at early and late times, $|\eta| \to \infty$, where
$|a'mk/\omega_k^3| \ll 1$.
This allows one to associate the Bunch-Davies vacuum with the choice $\tilde\alpha_k=1$
and $\tilde\beta_k= 0$ at early times $\eta \to -\infty$.
The vacua at early and late times, $|0\rangle_{\rm in}$ and $|0\rangle_{\rm out}$,
and the associated mode functions are related by the usual Bogolyubov coefficients
\be
u^{\rm in}_{\vec k,\lambda} = \alpha_k  u^{\rm out}_{\vec k,\lambda} + 
\beta_k  v^{\rm out}_{-\vec k,\lambda} ,
\ee
with the normalization $| \alpha_k |^2 + | \beta_k |^2 = 1$.
The Bogolyubov coefficient $\beta_k$ can be obtained from the mode functions by
\be
\beta_k = e^{i\lambda\Phi} \left( u_A^{\rm out} u_B^{\rm in} - u_B^{\rm out} u_A^{\rm in} \right) .
\ee
In terms of the coefficients $\tilde\alpha_k(\eta)$ and $\tilde\beta_k(\eta)$ we obtain
in the limit $\eta \to \infty$,
\be
| \beta_k | = \lim_{\eta \to \infty} | \tilde\beta_k(\eta) | ,
\ee
so that we only need to evaluate the late-time limit of $| \tilde\beta_k |$ to obtain the particle
production, as the energy density at late times reads as \cite{Chung:2011ck}
\be
\rho = \frac{4}{a^4} \int \frac{d\vec k}{(2\pi)^3} \omega_k | \beta_k |^2 ,
\ee
which gives Eq.(\ref{eq:rho-tilde-beta}).

\section{Gravitational production of Kaluza-Klein dark photons}
\label{app:production-photons}

In the appendix, we compute the gravitational production of the photons $A_\mu^{(c,n)}$ and
$A_\mu^{(s,n)}$ in the 4D Lagrangian (\ref{eq:L-4D-tot}), due to the expansion of the Universe.
This is similar to the computation presented in appendix~\ref{app:production-fermions}
for the Kaluza-Klein fermions.
We closely follow \cite{Kolb:2020fwh,Kolb:2023ydq}.
The Lagrangian (\ref{eq:L-Amu}) gives in curved spacetime the action
\be
\mathcal{S} = \int d^4x \, \sqrt{-g} \left[ - \frac{1}{4} g^{\mu\alpha} g^{\nu\beta} F_{\mu\nu} F_{\alpha\beta}
+ \frac{1}{2} m^2 g^{\mu\nu} A_\mu A_\nu  \right] ,
\label{eq:S-Amu}
\ee
with $m=n/R$ and
$F_{\mu\nu} = \nabla_\mu A_\nu - \nabla_\nu A_\mu = \partial_\mu A_\nu - \partial_\nu A_\mu$.
As in (\ref{eq:psi-modes-operators}), we can quantize the spin-1 field $A_\mu$ as
\ba
\hat A_\mu & = & \sum_\lambda \int \frac{d\vec k}{(2\pi)^{3/2}} \left[ \hat{a}_{\vec k,\lambda} 
u_{\vec k,\lambda,\mu}(\eta) e^{i\vec k \cdot \vec x} \right. \nonumber \\
&& \left. + \hat{a}_{\vec k,\lambda}^{\dagger} 
u_{\vec k,\lambda,\mu}^\star(\eta) e^{-i\vec k \cdot \vec x} \right] 
\label{eq:Amu-modes-operators}
\ea
where $\lambda=\pm 1,0$ is the helicity, $\hat a$ the ladder operators,
and $u_{\vec k,\lambda,\mu}(\eta)$ the mode functions.
They take the form
\be
u_{\vec k,\lambda,0}(\eta) = - i \frac{k_i u'_{\vec k, \lambda,i}}{k^2+a^2 m^2} ,
\ee
with
\be
u_{\vec k,\pm,i}(\eta) = \varphi_{k,\pm}(\eta) \epsilon_{\hat k, \pm, i} , 
\ee
for the transverse modes and
\be
u_{\vec k,0,i}(\eta) = \sqrt{\frac{k^2+a^2m^2}{a^2m^2}} \varphi_{k,0}(\eta) \epsilon_{\hat k, 0, i} 
\ee
for the longitudinal mode.
In spherical coordinates, with $\vec k = k (\sin\theta\cos\phi,\sin\theta\sin\phi,\cos\theta)$,
the polarization eigenvectors read as \cite{Kolb:2023ydq}
\be
\epsilon_{\hat{k},\pm} = \frac{i}{\sqrt{2}} e^{\mp i \gamma} \left( \begin{array}{c} \mp \cos\theta \cos\phi 
+ i \sin\phi \\  \mp \cos\theta - i \cos\phi  \\  \pm  \sin\theta \end{array} \right) , 
\ee
\be
\epsilon_{\hat{k},0} = \left( \begin{array}{c} \sin\theta \cos\phi \\ \sin\theta \sin\phi  \\  \cos\theta 
\end{array} \right) ,
\ee
with an irrelevant phase $\gamma$.
The mode functions obey the equations of motion \cite{Kolb:2023ydq}
\be
\varphi_{\lambda}'' + \omega_{\lambda}^2 \varphi_{\lambda} = 0 ,
\label{eq:motion-varphi}
\ee
with
\be
\omega_{\pm}^2 = k^2 + a^2 m^2 , 
\ee
and
\be
\omega_0^2 = k^2 + a^2 m^2 + \frac{k^2}{6} \frac{a^2 R}{k^2+a^2 m^2} 
+ 3 k^2 \frac{a^4 H^2 m^2}{(k^2+a^2m^2)^2} ,
\ee
where $R$ is the Ricci scalar.
The creation and annihilation operators obey the commutation relations
\be
[ \hat{a}_{\vec k,\lambda} , \hat{a}_{\vec k',\lambda'}^\dagger ] = 
\delta_D(\vec k - \vec k') \delta_{\lambda,\lambda'} ,
\ee
while the mode functions satisfy the normalization
\be
\varphi'^\star \varphi - \varphi^\star \varphi' = i .
\ee
The vacua at early and late times, $|0\rangle_{\rm in}$ and $|0\rangle_{\rm out}$,
and the associated mode functions are related by the usual Bogolyubov coefficients
\be
\varphi^{\rm in}_{k} = \alpha_k  \varphi^{\rm out}_{k} + \beta_k  \varphi^{\rm out \star}_{k} ,
\label{eq:photons-in-out}
\ee
with the normalization $| \alpha_k |^2 - | \beta_k |^2 = 1$.
The Bogolyubov coefficient $\beta_k$ can be obtained from the mode functions by
\be
\beta_k = i \left( \varphi_k^{\rm out} \,\! ' \varphi_k^{\rm in} - \varphi_k^{\rm out} \varphi_k^{\rm in} \,\! '
\right) .
\label{eq:betak-modes}
\ee
Again, we use the adiabatic representation \cite{Kofman:1997yn} by writing the mode functions
in terms of the instantaneous vacuum mode functions,
\be
\varphi_k(\eta) = \tilde\alpha_k(\eta) \frac{e^{-i\Phi_k}}{\sqrt{2\omega_k}} + 
\tilde\beta_k(\eta) \frac{e^{i\Phi_k}}{\sqrt{2\omega_k}} ,
\label{eq:photons-adiabatic}
\ee
where we impose the constraint
\be
\varphi_k'(\eta) = - i \omega_k \tilde\alpha_k(\eta) \frac{e^{-i\Phi_k}}{\sqrt{2\omega_k}} + 
i \omega_k \tilde\beta_k(\eta) \frac{e^{i\Phi_k}}{\sqrt{2\omega_k}} ,
\ee
and we introduced
\be
\Phi_k(\eta) = \int^\eta d\eta' \, \omega_k(\eta') .
\ee
Then, the equation of motion (\ref{eq:motion-varphi}) gives the time evolution
\be
\tilde\alpha_k ' = \tilde\beta_k \frac{\omega_k'}{2\omega_k} e^{2 i \Phi_k} ,\;\;
\tilde\beta_k ' = \tilde\alpha_k \frac{\omega_k'}{2\omega_k} e^{-2 i \Phi_k} .
\label{eq:motion-adiabatic-photons}
\ee
The coefficients $\tilde\alpha_k$ and $\tilde\beta_k$ satisfy the normalization
$| \tilde\alpha_k |^2 - | \tilde\beta_k |^2 = 1$. 
At early times, $\eta \to -\infty$, we have $\tilde\alpha_k=1$ and $\tilde\beta_k=0$
by definition of the ``in'' vacuum, and from Eq.(\ref{eq:betak-modes})
the coefficient $\beta_k$ is given by
\be
\beta_k = \lim_{\eta\to\infty} \tilde\beta_k(\eta) ,
\label{eq:tilde-beta-photons}
\ee
so that for our purpose it is sufficient to compute the value of $\tilde\beta_k$ at late times.
The energy density then reads as
\be
\rho = \frac{1}{a^4} \int \frac{d\vec k}{(2\pi)^3} \omega_k | \beta_k |^2 ,
\label{eq:rho-photons}
\ee
for a given polarization.

\bibliography{ref}

\end{document}